\documentclass{article}
\usepackage{graphicx, amsmath, tikz, float, comment, array, multirow, booktabs, makecell, tabularx, ragged2e, authblk, setspace, hanging,tabularx, lineno, geometry, bbding, subfig, bbold, dsfont, algorithm2e, indentfirst} 
\usepackage[hidelinks]{hyperref}
\usepackage{listings}
\usepackage{courier}
\usepackage{graphicx}
\usepackage[bb=px]{mathalfa} 
\usepackage{color}
\usepackage{geometry}[margin=2in]
\usepackage{natbib}
\usepackage[makeroom]{cancel}

\RestyleAlgo{ruled}
\SetKwInput{KwParams}{Parameters}
\SetKwInput{KwKnown}{Known}
\SetKwBlock{FirstStage}{First Stage:}{\ensuremath{\text{Output: $\big\{ \boldsymbol{\beta}^{(k)} | \{ \mathbf{s}_i \}, n \big\}_{k = 1}^K$ }}}
\SetKwBlock{InterStage}{Intermediate Stage:}{\ensuremath{\text{Output: $\big\{ \Lambda^{(k)}(\mathcal{S}) \big\}_{k = 1}^K$}}}
\SetKwBlock{SecondStage}{Second Stage:}{\ensuremath{\text{Output: $\big\{ {\beta_0}^{(k)}, \boldsymbol{\beta}^{(k)} | \{ \mathbf{s}_i \}, n \big\}_{k = 1}^K$ }}}
\SetKwBlock{DoParallel}{do in parallel}{end}

\newcolumntype{C}{ >{ \arraybackslash \Centering } X }

\title{A multi-stage Bayesian approach to fit spatial point process models}
\author[1,*]{Rachael Ren}
\author[1]{Mevin B. Hooten}
\author[2]{Toryn L.J. Schafer}
\author[1]{Nicholas M. Calzada}
\author[2]{Benjamin Hoose}
\author[3]{Jamie N. Womble}
\author[3]{Scott Gende}

\affil[1]{Department of Statistics and Data Sciences, The University of Texas at Austin, Austin, TX, USA}
\affil[2]{Department of Statistics, Texas A\&M University, College Station, TX, USA}
\affil[3]{Glacier Bay National Park
and Preserve and Southeast Alaska Network, National Park Service, Juneau, AK, USA}
\affil[*]{\textit{email:} \href{mailto:rren@utexas.edu}{rren@utexas.edu}}
\date{}

\begin{document}

\maketitle

\setstretch{2}

\section*{Abstract}

\noindent Spatial point process (SPP) models are commonly used to analyze point pattern data in many fields, including presence-only data in ecology. Existing exact Bayesian methods for fitting these models are computationally expensive because they require approximating an intractable integral each time parameters are updated and often involve algorithm supervision (i.e., tuning in the Bayesian setting). We propose a flexible, efficient, and exact multi-stage recursive Bayesian approach to fitting SPP models that leverages parallel computing resources to obtain realizations from the joint posterior, which can then be used to obtain inference on derived quantities. We outline potential extensions, including a framework for analyzing study designs with compact observation windows and a neural network basis expansion for increased model flexibility. We demonstrate this approach and its extensions using a simulation study and analyze data from aerial imagery surveys to improve our understanding of spatially explicit abundance of harbor seal (\textit{Phoca vitulina}) pups in Johns Hopkins Inlet, a protected tidewater glacial fjord in Glacier Bay National Park, Alaska.

\section{Introduction}

Spatial point process (SPP) models are a class of generative stochastic models that give rise to a random number of irregularly spaced events in a spatial domain. They have been used to analyze point patterns in various fields, such as ecology \citep[e.g.,][]{Haase1995, Law2009, WartonShepherd2010, Renner2015}, seismology \citep[e.g.,][]{Ogata1998}, and social sciences \citep[e.g.,][]{Mohler2011}. Many models have been developed for SPP data including, but not limited to, the homogeneous and inhomogeneous Poisson point process, the log-Gaussian Cox process \citep[LGCP;][]{Moller1998}, and the Hawkes process \citep{Hawkes1971}. This article focuses on a computational method for fitting the inhomogeneous Poisson Process (IPP), a point process model in which point intensity varies across space, and extensions that allow for increased model flexibility.

The stochasticity in SPP data arises from two sources: the random number of point events and their locations, which we refer to as $n$ and $\{\mathbf{s}_i\}_{i = 1}^n$, respectively. Although it is common in the species distribution modeling (SDM) literature to fit an IPP model using a likelihood conditioned on $n$ \citep[e.g.,][]{Johnson2013}, we focus on a complete IPP likelihood that models both data types jointly, allowing spatial heterogeneity and total abundance (i.e., the total number of points in the study domain) to be modeled in a unified framework. This feature is especially useful when the data are partially observed (i.e., total abundance is unknown and may exceed $n$) and inference on total abundance and/or prediction in unobserved areas is desired. We extend the complete IPP likelihood to data observed in compact observation windows, a special case of partially observed data defined by \citet{Baddeley2015} that may arise naturally from certain study designs, such as aerial imagery surveys. 

IPP likelihoods are often intensive to compute because they involve an integral over the spatial domain, which is analytically intractable and must be approximated using methods such as numerical quadrature. This is especially inefficient when using Monte Carlo Markov Chain \citep[MCMC;][]{GelfandSmith1990} methods because the integral is necessarily approximated each time the parameters are updated. Furthermore, conventional MCMC algorithms for fitting IPP models often require algorithmic tuning due to a lack of conjugate prior distributions. Various fast approximate Bayesian methods for fitting SPP models, such as the Integrated Nested Laplace Approximation \citep[INLA;][]{Rue2009} approach to fitting LGCP models \citep{Illian2012}, circumvent this issue by approximating marginal posterior distributions. However, obtaining exact Bayesian inference for the joint posterior distribution remains computationally challenging. Obtaining realizations from the joint posterior distribution, as opposed to approximating marginal distributions, may be beneficial for estimating derived quantities of interest, such as total abundance.  

One alternative to fitting an exact Bayesian model using conventional MCMC is to use recursive Bayes approaches, also referred to as Bayesian filtering or sequential inference, to fit the model in stages \citep{Särkkä2013}. Although these approaches are widely applicable, recursive Bayes approaches for fitting various ecological models have been developed in recent years due to their natural application to hierarchical models and computational efficiency \citep[e.g.,][]{Hooten2023, Hooten2024, Johnson2022, Leach2022, McCaslin2020}. These methods often implement a version of the prior-proposal recursive Bayes (PPRB) method described by \citet{Hooten2021}. In a two-stage implementation of PPRB, the data are partitioned into two subsets, $\mathbf{y} := (\mathbf{y}_1, \mathbf{y}_2)^\prime$. We let $\boldsymbol{\theta}$ represent a vector of model parameters and $[\boldsymbol{\theta}]$ represent its prior distribution. We use the bracket notation `$[\cdot]$' to represent probability distributions hereafter \citep{GelfandSmith1990}. In the first stage, we use $\mathbf{y}_1$ to find a transient posterior distribution $[\boldsymbol{\theta} | \mathbf{y}_1] \propto [\mathbf{y}_1 | \boldsymbol{\theta}] [\boldsymbol{\theta}]$. Next, in the second stage, we use the transient posterior distribution as a prior and proposal distribution in an MCMC algorithm to obtain a sample from the full posterior distribution
\begin{align*}
    [\boldsymbol{\theta} | \mathbf{y}] &\propto [\mathbf{y} | \boldsymbol{\theta}] [\boldsymbol{\theta}]\\
    &\propto [\mathbf{y}_2 | \boldsymbol{\theta}, \mathbf{y}_1][\boldsymbol{\theta} | \mathbf{y}_1].
\end{align*}
Critically, using the transient posterior distribution as both the prior and proposal distribution results in cancellations in the Metropolis-Hastings (M-H) ratio for the second stage. Furthermore, the multi-stage implementation allows computationally intensive calculations involved in evaluating the conditional likelihood $[\mathbf{y}_2 | \boldsymbol{\theta}, \mathbf{y}_1]$ to be computed in parallel in an intermediate stage.

Conventionally, the partitions in PPRB comprise data of the same type, but alternative perspectives on partitioning can be helpful in some cases. For instance, PPRB approaches for fitting capture-recapture models partition the data such that the first partition comprises the observed binary capture histories and the second partition comprises the number of observed individuals \citep{Hooten2023, Hooten2024}. Similarly, we show that it is natural to partition spatial point pattern data such that the first partition comprises the observed locations $\{ \mathbf{s}_i \}_{i = 1}^n$ and the second partition comprises the number of observed point events $n$. We can then fit the data model conditioned on $n$ in the first stage and assimilate $n$ in the second stage. This partitioning scheme allows us to take advantage of well-known strategies that use the conditional data model $[\{ \mathbf{s}_i\} | \boldsymbol{\theta}, n]$, such as the logistic and Poisson regression approximations \citep{Aarts2012, Baddeley2010, WartonShepherd2010}. We propose a multi-stage MCMC method for fitting the IPP model using PPRB, along with a variety of strategies for first-stage sampling from the transient posterior distribution, resulting in an efficient and flexible approach to fitting an exact Bayesian SPP model.

In section 2, we provide an overview for fitting an IPP model and potential extensions for additional model flexibility. Next, in section 3, we construct a Bayesian IPP model, propose a recursive Bayes procedure for fitting the model, and compare various first-stage sampling strategies. We then demonstrate the multi-stage approach on simulated data in a compact window setting in section 4. Finally, in section 5, we apply our approach to data from aerial imagery surveys to learn spatially explicit abundance of harbor seals in Johns Hopkins Inlet, a protected glacial fjord in Glacier Bay National Park, Alaska, and describe strategies to account for additional spatial heterogeneity in the model.

\section{Inhomogeneous Poisson Process}

We consider a space of interest, $\mathcal{S} \subset \mathbb{R}^2$, and a set of disjoint subspaces $\{A_l\}_{l = 1}^L$ such that $\mathcal{S} = \cup_l A_l$. We let $\mathbf{s}_{i} \in \mathcal{S}$ be a two-dimensional vector denoting event location for $i = 1, \dots, n$ observed events. We also denote $N(\cdot)$ as a random counting measure and $N$ as the total number of events in $\mathcal{S}$ (i.e., $N = N(\mathcal{S})).$

For an inhomogeneous Poisson process,
\begin{enumerate}
    \item[i)] $N(A_l)$ follows a Poisson distribution with rate parameter $\Lambda(A_l) := \int_{A_l} \lambda(\mathbf{s})\text{d}\mathbf{s}$.
    \item[ii)] $N(A_1), \dots, N(A_L)$ are conditionally independent random variables for $l = 1, \dots, L$.
\end{enumerate} 
An IPP can also be interpreted as the continuous limit of an independent Poisson count model for an increasingly fine grid of subspaces (i.e., as $L$ approaches $\infty$) over the spatial domain \citep{FithianHastie2013}. Additional details and theoretical properties on the IPP are available in \citet{Illian2008} and \citet{Gelfand2010}.

For $p$ covariates associated with each point in $\mathcal{S}$, we can model the IPP intensity $\lambda$ as a function of the spatially referenced covariates
$\lambda(\mathbf{s}) = f(\beta_0 + \mathbf{x}^\prime(\mathbf{s})\boldsymbol{\beta}\bigr)$, where $\beta_0$ represents the intercept, $\mathbf{x}$ and $\boldsymbol{\beta}$ denote $p \times 1$ vectors of covariates and their corresponding coefficients, respectively, and the function $f$ represents a positive strictly increasing differentiable function. We assume the convention $f(x) = \exp(x)$. If desired, this assumption may be relaxed by using a different link function and/or transforming the covariates using a basis expansion. We discuss the latter in section 5.

\subsection{Complete Likelihood}

The IPP likelihood that jointly models the number of observed points and their locations is as follows
\begin{equation}
    \big[\{ \mathbf{s}_{i}\}, n | \beta_0,  \boldsymbol{\beta}\big] = \frac{\prod_{i = 1}^{n} \lambda(\mathbf{s}_{i})}{n! \exp\bigl( \Lambda(\mathbf{\mathcal{S}}) \bigr)},
\end{equation}
\noindent for $i = 1, \dots, n$. We refer to this IPP likelihood in (1) as the complete likelihood hereafter. 

In practice, the integral $\Lambda(\mathcal{S})$ in (1) is not analytically tractable and must be approximated. Using covariate information for grid centers $\mathbf{u}_l$ for $l = 1, \dots, L$, we can approximate $\Lambda(\mathcal{S})$ with numerical quadrature and use $\{ \mathbf{u}_l \}_{l = 1}^L$ as quadrature points, but this can be computationally intensive, especially for large $L$. 

\subsection{Conditional Likelihood}

The IPP likelihood conditional on $n$ is as follows
\begin{equation}
    \big[\{ \mathbf{s}_i\}  | n, \boldsymbol{\beta}\big] = \frac{\prod_{i = 1}^n \lambda(\mathbf{s}_i)}{\bigl( \Lambda(\mathcal{S}) \bigr)^n},
\end{equation}
for $i = 1, \dots, n$. We refer to this as the conditional likelihood hereafter. It is important to note that unlike in the complete likelihood, $\beta_0$ is non-identifiable because it cancels in the numerator and denominator in (2); however, inference on $\beta_0$ is necessary to estimate total abundance for partially observed domains. We let $\boldsymbol{\beta}$ denote the vector of coefficient values excluding the intercept hereafter for clarity.

The conditional likelihood (2) is commonly used in SDM literature and is appropriate for applications where the number of observed points $n$ is treated as known and fixed \textit{a priori}. For instance, in telemetric surveys for studying animal movement \citep{Hooten2017}, it is common to fit an SPP model to the observed spatially explicit points for a predetermined number of animal relocations \citep[e.g.,][]{Johnson2013}. The conditional likelihood suffers from similar computational challenges in approximating $\Lambda(\mathcal{S})$; however, useful likelihood approximations have been developed to circumvent this issue.

\subsubsection{Logistic Regression Approximation}
One widely adopted approximation for the conditional likelihood (2) in both point process and SDM literature is the logistic regression approximation, which involves augmenting $m$ background points, also referred to as pseudo-absences, to the data to represent locations where point events were absent \citep{Baddeley2010, WartonShepherd2010, FithianHastie2013}. \citet{WartonShepherd2010} showed that if the number of observed points $n$ is fixed and the number of background points $m$ grows infinitely large, then the logistic regression coefficients (intercept excluded) converge to the maximum likelihood estimates of the IPP coefficients in the conditional likelihood (2).  

To implement this approach, we use the Berman-Turner device \citep{BermanTurner1992}, where we define an indicator variable, $y(\mathbf{s}) = \mathds{1}(\mathbf{s} \in \{ \mathbf{s}_i\}_{i = 1}^n)$, which indicates whether a location $\mathbf{s}$ is within our set of observed points. By definition, we have $y(\mathbf{s}_i) = 1$ for $i = 1, \dots, n$. To obtain points where $y(\mathbf{s}) = 0$, we generate a uniform random sample of size $m$ within $\mathcal{S}$ (see \citet{Northrup2013} for guidance on choosing the number of background points $m$ in practice). For $m$ background points, we order $y(\mathbf{s}_i)$ such that $y(\mathbf{s}_i) = 1$ for $i = 1, \dots, n$ and $y(\mathbf{s}_i) = 0$ for $i = n+1, \dots, \tilde{n}$ where $\tilde{n} = n + m$. We then fit $y(\mathbf{s}_i) \sim \text{Bern}(p_i)$ where $\text{logit}(p_i) = \beta_0 + \mathbf{x}^\prime (\mathbf{s}_i)\boldsymbol{\beta}.$ Although the logistic regression intercept is not relevant for the IPP model, note that its inclusion is necessary to adjust for the background sample size \citep{Fieberg2021}. 

\subsection{Compact Observation Window Extension}

In some scenarios, point events are only observed within compact observation windows. To extend the likelihood to a compact window setting, we denote each window using $\mathcal{S}_j$ for $j = 1, \dots, J$. The remaining unobserved area is defined as $\mathcal{S}_0 := \mathcal{S} \backslash \cup_j \mathcal{S}_j$ (see Figure 2b for a visual representation of a compact window setting). We focus on the case where the probability of detecting point events in $\cup_j \mathcal{S}_j$ is 1 (i.e., perfect detection) and the probability of detecting point events in $\mathcal{S}_0$ is 0 (i.e., no detection). In such cases, we are often interested in inference on $N$, the total number of points in $\mathcal{S}$.

The joint likelihood of $\{ \mathbf{s}_{ij} \}$, the set of event locations for each individual $i$ in the $j$th window, and $\{n_j\}$, the set of point event counts within each window $j$, becomes
\begin{equation}
    \big[ \{ \mathbf{s}_{ij}\}, \{ n_j\} | \beta_0,  \boldsymbol{\beta}\big] = \frac{\prod_{\{ \forall j, n_j > 0\}}\prod_{i = 1}^{n_j} \lambda(\mathbf{s}_{ij})}{n! \exp\bigl( \Lambda(\cup_j \mathcal{S}_j) \bigr)},
\end{equation}
for $i = 1, \dots, n$ and $j = 1, \dots, J$ where $n = \sum_j n_j$. We refer to (3) as the windowed complete likelihood hereafter. Similar to the conditional likelihood (2), the windowed conditional likelihood can be derived by conditioning on $\{n_j\}.$

\section{Bayesian Model}

Using the complete likelihood (1), we can estimate and quantify uncertainty for the parameters of interest, $\beta_0$ and $\boldsymbol{\beta}$, using a Bayesian model. The full posterior distribution associated with our model is as follows:
\begin{equation}
    \big[\beta_0, \boldsymbol{\beta}| \{\mathbf{s}_{i}\}, n\big] \propto \big[ \{ \mathbf{s}_{i}\}, n | \beta_0,  \boldsymbol{\beta}\big] [\beta_0] [\boldsymbol{\beta}].
\end{equation}

Recall that, for the complete likelihood (1), $\Lambda(\mathcal{S})$ is analytically intractable. Using numerical quadrature to approximate this integral often involves computing a large sum that scales with the number of quadrature points. Furthermore, because $\Lambda(\mathcal{S})$ depends on $\beta_0$ and $\boldsymbol{\beta}$, the integral must be approximated each time the parameters are updated.

Secondly, we are unable to obtain closed form representations for the full-conditional distributions for $\beta_0$ and $\boldsymbol{\beta}$. If we use a normal random walk proposal for $\beta_0$ and $\boldsymbol{\beta}$, we must perform tuning, which incurs additional computational costs. Unsupervised MCMC algorithms, such as Hamiltonian Monte Carlo \citep[HMC;][]{Neal2011} with the No-U-Turn Sampler \citep[NUTS;][]{HoffmanGleman2014}, may also be implemented, but require overhead calculations.

Instead of evaluating the full posterior distribution (4) using a conventional single-stage MCMC algorithm, we propose a multi-stage MCMC algorithm for fitting the model using the complete likelihood that does not require parameter tuning and allows for the $\Lambda(\mathcal{S})$ approximation to be computed in parallel.

\subsection{Recursive Bayes Algorithm}

To construct our multi-stage MCMC algorithm, we apply PPRB following \citet{Hooten2021}. Our algorithm follows a similar structure of the two-stage PPRB algorithm summarized in section 1 with the addition of an intermediate stage for parallel computation.

We first show that the complete likelihood (1) can be decomposed into two terms:
\begin{align}
    \big[ \{\mathbf{s}_{i}\}, n | \beta_0\ \boldsymbol{\beta}\big] &= \frac{\prod_{i = 1}^{n} \lambda(\mathbf{s}_{i})}{n! \exp\bigl(\Lambda(\mathcal{S})\bigr)}\\
    &= \Biggl( \frac{\prod_{i=1}^n \lambda(\mathbf{s}_i)}{ \big( \Lambda(\mathcal{S}) \big)^n} \Biggr) \Biggl( \frac{\big( \Lambda(\mathcal{S}) \big)^n }{n! \exp\big(\Lambda(\mathcal{S}) \big)}\Biggr)\\
    &= \big[\{ \mathbf{s}_{i} \} | \boldsymbol{\beta}, n\big] [ n | \beta_0, \boldsymbol{\beta}].
\end{align}
Moreover, note that the first term in (7) is equivalent to the conditional likelihood (2). The second term, the likelihood for $n$, follows a Poisson distribution:
\begin{equation}
    [n | \beta_0, \boldsymbol{\beta}] = \text{Pois}\bigl(\Lambda(\mathcal{S})\bigr).
\end{equation}
The full posterior distribution can then be decomposed as 
\begin{equation}
    \big[\beta_0, \boldsymbol{\beta}| \{\mathbf{s}_{i}\}, n\big] \propto \big[\{ \mathbf{s}_{i} \} | \boldsymbol{\beta}, n\big] [ n | \beta_0, \boldsymbol{\beta}] [\beta_0] [\boldsymbol{\beta}].
\end{equation}

Evaluating the decomposed posterior distribution (9) using PPRB results in two stages. In the first stage, we evaluate the transient posterior using the conditional likelihood (2) and the prior for $\boldsymbol{\beta}$
\begin{equation}
    \big[ \boldsymbol{\beta} | \{ \mathbf{s}_i\}, n\big] \propto \big[\{ \mathbf{s}_{i} \} | \boldsymbol{\beta}, n\big] [\boldsymbol{\beta}]
\end{equation}
to obtain an MCMC sample, $\big\{ \boldsymbol{\beta}^{(k)} | \{ \mathbf{s}_i \}, n \big\}_{k = 1}^K$. Recall that in section 2.2.1, we show that fitting the model using the conditional likelihood (2) can be approximated using logistic regression, which permits a variety of ways to obtain a sample from the transient posterior distribution. We describe and compare first-stage sampling strategies in the context of the multi-stage algorithm in the following section.

In the intermediate stage, we use the transient posterior realizations to compute corresponding values of $\Lambda(\mathcal{S})$ in parallel and store them for recall in the following stage. Approximating $\Lambda(\mathcal{S})$ is the most computationally intensive step in evaluating the likelihood for $n$ and parallelization results in significant speedup for the final stage of the algorithm.

In the second and final stage, we use the transient posterior distribution as a prior and proposal distribution in an MCMC algorithm to account for the stochasticity associated with $n$ in the full posterior distribution (4). Using the transient posterior distribution (10) as a prior and proposal distribution for $\boldsymbol{\beta}$ results in convenient cancellations involving the data model conditioned on $n$ in the M-H acceptance ratio: 
\begin{align}
    \alpha &= 
    \frac{\big[ \{\mathbf{s}_i\}, n | {\beta_0}^{(*)}, \boldsymbol{\beta}^{(*)} \big] 
    \big[ {\beta_0}^{(*)} \big] 
    \big[ {\beta_0}^{(k-1)} \big]_*
    \big[ {\boldsymbol{\beta}}^{(*)} \big] 
    \big[ {\boldsymbol{\beta}}^{(k-1)} \big]_*}
    {\big[ \{\mathbf{s}_i\}, n | {\beta_0}^{(k-1)}, \boldsymbol{\beta}^{(k-1)} \big] 
    \big[ {\beta_0}^{(k-1)} \big] 
    \big[ {\beta_0}^{(*)} \big]_*
    \big[ {\boldsymbol{\beta}}^{(k-1)} \big] 
    \big[ {\boldsymbol{\beta}}^{(*)} \big]_*}\\[1em]
    &=
    \frac{\big[ \{\mathbf{s}_i\} | {\beta_0}^{(*)}, \boldsymbol{\beta}^{(*)} \big] 
    \big[ n | {\beta_0}^{(*)}, \boldsymbol{\beta}^{(*)} \big]
    \big[ {\beta_0}^{(*)} \big] 
    \big[ {\beta_0}^{(k-1)} \big]_*
    \big[ {\boldsymbol{\beta}}^{(*)} \big] 
    \big[ {\boldsymbol{\beta}}^{(k-1)} \big]_*}
    {\big[ \{\mathbf{s}_i\} | {\beta_0}^{(k-1)}, \boldsymbol{\beta}^{(k-1)} \big] 
    \big[ n | {\beta_0}^{(k-1)}, \boldsymbol{\beta}^{(k-1)} \big]
    \big[ {\beta_0}^{(k-1)} \big] 
    \big[ {\beta_0}^{(*)} \big]_*
    \big[ {\boldsymbol{\beta}}^{(k-1)} \big] 
    \big[ {\boldsymbol{\beta}}^{(*)} \big]_*}\\[1em]
    &=
    \frac{\bigl[ n | {\beta_0}^{(*)}, \boldsymbol{\beta}^{(*)} \bigr] 
    \big[ {\beta_0}^{(*)} \big] 
    \big[ {\beta_0}^{(k-1)} \big]_*}
    {\bigl[ n | {\beta_0}^{(k-1)}, \boldsymbol{\beta}^{(k-1)} \bigr]
    \big[ {\beta_0}^{(k-1)} \big] 
    \big[ {\beta_0}^{(*)} \big]_*}.
\end{align}
The $(*)$ superscript denotes a proposed value, the $(k-1)$ superscript denotes the value of a parameter at the $(k-1)$th step of the MCMC chain, and $[\cdot]_{*}$ denotes a proposal distribution. Critically, we do not need to perform additional calculations using the point data $\{\mathbf{s}_i\}$ at this stage.

Recall that we are unable to learn about $\beta_0$ using the first-stage logistic regression approximation. It is, however, crucial to efficiently propose $\beta_0$ for the M-H correction in the second stage to obtain reasonable MCMC mixing rates and prevent sample degeneracy. Conveniently, if we transform the intercept such that $\zeta = e^{\beta_0}$, we can obtain a Gibbs sampler using a $\text{Gamma}(a,b)$ prior for $\zeta$. The Gibbs update for $\zeta$ is based on the full-conditional distribution 
\begin{equation}
    [\zeta | \cdot ] = \text{Gamma}\big(a + n, b + \Lambda(\mathcal{S}) \big).
\end{equation}
We then log-transform our posterior sample for $\zeta$ to obtain a sample from the full-conditional distribution $[\beta_0 | \cdot]$. For a full derivation of the Gibbs update for $\zeta$, see Appendix A.

The Gibbs sampler for $\beta_0$ also allows for the M-H ratio (13) to simplify to
\begin{equation}
    \alpha = \big[ n | {\beta_0}^{(k)}, \boldsymbol{\beta}^{(*)}\big] / \big[ n | {\beta_0}^{(k)},  \boldsymbol{\beta}^{(k-1)}\big].
\end{equation} Thus, each MCMC iteration in the second stage only requires updating $\beta_0$ using the Gibbs update (14) and evaluating the Poisson likelihood $\big[ n | {\beta_0}^{(k)}, \boldsymbol{\beta}^{(*)}\big]$, both of which are fast using the pre-computed values of $\Lambda(\mathcal{S})$ from the intermediate stage. The resulting MCMC realizations from the second stage are realizations from the full posterior distribution (4). The full multi-stage algorithm is summarized in Algorithm 1.

\begin{algorithm}
\DontPrintSemicolon
\SetAlgoLined

\FirstStage{
\KwData{$\{s_i\}_{i=1}^n$}
\KwParams{$\boldsymbol{\beta}$}
\For{$k = 1, \dots, K$}{
    Sample from $\big[ \boldsymbol{\beta} | \{ \mathbf{s}_i\}, n\big] \propto \big[\{ \mathbf{s}_{i} \} | \boldsymbol{\beta}, n\big] [\boldsymbol{\beta}].$
}
}

\vspace{1em}
\InterStage{
\KwIn{$\big\{ \boldsymbol{\beta}^{(k)} | \{ \mathbf{s}_i \}, n \big\}_{k = 1}^K$}
\DoParallel{
    \For{$k = 1, \dots, K$}{
        Approximate $\Lambda^{(k)}(\mathcal{S})$ using first-stage $\boldsymbol\beta^{(k)}.$
    }
}
}

\vspace{1em}
\SecondStage{
\KwData{$n$}
\KwParams{$\beta_0, \, \boldsymbol{\beta}$}
Initialize ${\beta_0}$, ${\boldsymbol{\beta}}$ and compute corresponding $\Lambda(\mathcal{S})$. \;
\For{$k = 1, \dots, K$}{
    Sample  $\zeta^{(k)} \sim \text{Gamma}\big(a + n, b + \Lambda^{(k)}(\mathcal{S}) \big)$.\;
    Set ${\beta_0}^{(k)} = \log\big(\zeta^{(k)}\big).$\;
    Sample uniformly with replacement $\boldsymbol{\beta}^{(*)} \sim \big\{ \boldsymbol{\beta}^{(k)} | \{ \mathbf{s}_i \}, n \big\}_{k = 1}^K$. \;
    Fetch corresponding $\Lambda^{(*)}(\mathcal{S})$. \;
    Set $\boldsymbol{\beta} = \boldsymbol{\beta}^{(*)}$ and $\Lambda(\mathcal{S}) = \Lambda^{(*)}(\mathcal{S})$ w.p. $\alpha = \big[ n | {\beta_0}^{(k)}, \boldsymbol{\beta}^{(*)}\big] / \big[ n | {\beta_0}^{(k)},  \boldsymbol{\beta}^{(k-1)}\big]$. \;
    Set $\boldsymbol{\beta}^{(k)} = \boldsymbol{\beta}$ and $\Lambda^{(k)}(\mathcal{S}) = \Lambda(\mathcal{S}).$ \;
}
}

\caption{Multi-stage PPRB MCMC}
\end{algorithm}

\subsection{First-stage Sampling Strategies}

One advantage to fitting the Bayesian SPP model in multiple stages is the increased flexibility facilitated by the ability to approximate the first stage using a logistic regression model. There are a variety of approaches for fitting a Bayesian logistic regression model such as the Pólya-Gamma data augmentation method described by \citet{Polson2013}, which results in a fully Gibbs algorithm and alleviates the need to perform parameter tuning. Other unsupervised Monte Carlo methods, such as HMC with NUTS, for fitting logistic regression can be performed using existing software, such as \texttt{rstanarm} \citep{rstanarm} or \texttt{brms} \citep{brms} in \texttt{R}. We refer to fitting the multi-stage procedure with the Pólya-Gamma data augmentation and HMC approaches as the Pólya-Gamma (PG) and HMC methods, respectively, in reference to their first-stage sampling strategies.

An alternative to fitting the multi-stage algorithm as fully Bayesian is to fit the first stage using maximum likelihood estimation. This can be attractive because fitting a Bayesian logistic regression model in the first stage can be slow, especially after performing large data augmentations, such as the background points generated from the Berman-Turner device and/or the latent variables required for the Pólya-Gamma Gibbs sampler. This creates a computational bottleneck because the first stage cannot be parallelized and thus does not scale with the number of computational cores. \citet{Johnson2022} proposed a multi-stage PPRB algorithm for a different class of models in which the transient posterior distribution is approximated using non-Bayesian likelihood maximization in the first stage. A sample from the approximated transient posterior distribution can then be corrected using an approximate M-H ratio in later stages \citep{Johnson2022}. 

Applying a similar approach to fit SPP models, we can perform fast maximum likelihood estimation for $\boldsymbol{\beta}$ in the first stage using the \texttt{glm} function in \texttt{R} to fit the logistic regression model described in section 2.2.1. Assuming a flat prior for $\boldsymbol{\beta}$ and standard regularity conditions (i.e., the likelihood is continuously twice-differentiable), we can apply large sample theory to obtain the following approximation: 
\begin{equation}
    \Big[ \boldsymbol{\beta} | \big\{ y(\mathbf{s}_i) \big\}, n \Big] \approx \text{N}\big( \hat{\boldsymbol{\beta}}, \hat{\boldsymbol{\Sigma}}_\beta \big),
\end{equation}
where $\hat{\boldsymbol{\beta}}$ and $\hat{\boldsymbol{\Sigma}}_\beta$ are the maximum likelihood estimates and inverse observed Fisher information, respectively. We assume that the sample size is large enough for $\hat{\boldsymbol{\beta}}$ to be identifiable and well-defined. Next, we draw an i.i.d. sample from the approximate Gaussian distribution in (16) as an approximate sample from the transient posterior distribution. 

\citet{Johnson2022} showed that the approximate transient posterior distribution (16) allows for approximate cancellations in the later stage M-H ratio and that parameter estimates obtained using this method are close to those obtained through a conventional single-stage MCMC algorithm. This method can be applied in our multi-stage procedure to obtain an approximate and fast sample from the posterior distribution for $\boldsymbol{\beta}$. Using the Gibbs sampler in (14), the M-H ratio in the second stage again reduces to (15). We refer to this multi-stage method as the GLM approximate (GLM-A) method hereafter.

\subsubsection{GLM Exact Approach}

If exact inference is desired, we propose an adjustment to the approximate approach developed by \citet{Johnson2022} to make it exact. After performing maximum likelihood estimation and obtaining an approximate transient posterior distribution as in (16), we can obtain an exact M-H correction for the complete likelihood (1) that retains the contribution from the original prior for $\boldsymbol{\beta}$ and the approximate transient posterior proposal distribution. Using the Gibbs sampler for $\zeta$ based on the full-conditional distribution in (14) results in the following M-H ratio: 
\begin{equation}
     \alpha = \frac{\big[ \{\mathbf{s}_i\}, n | {\beta_0}^{(*)}, \boldsymbol{\beta}^{(*)} \big] 
    \big[ {\boldsymbol{\beta}}^{(*)} \big] 
    \big[ {\boldsymbol{\beta}}^{(k-1)} \big]_*}
    {\big[ \{\mathbf{s}_i\}, n | {\beta_0}^{(k-1)}, \boldsymbol{\beta}^{(k-1)} \big] 
    \big[ {\boldsymbol{\beta}}^{(k-1)} \big] 
    \big[ {\boldsymbol{\beta}}^{(*)} \big]_*}.
\end{equation}
Note that this approach differs from the previously described PPRB strategies (PG, HMC, and GLM-A) because we use the naive (i.e., noninformative) prior as opposed to the transient posterior distribution as the prior distribution for $\boldsymbol{\beta}$. However, this approach still differs from a traditional MCMC algorithm in that it uses a PPRB-style strategy to obtain an efficient proposal distribution and a multi-stage construction to leverage parallel computing resources.

To accelerate the evaluation of the log-likelihood for $\big[ \{\mathbf{s}_i\}, n | {\beta_0}^{(*)}, \boldsymbol{\beta}^{(*)} \big]$ in the second stage, we compute $\sum_{i = 1}^n \mathbf{x}^\prime(\mathbf{s}_i)\boldsymbol{\beta}$ in parallel in the intermediate stage in addition to $\Lambda(\mathcal{S})$ using the first stage transient posterior sample. Although the intermediate and second stages involve more calculations compared to the previous multi-stage methods (PG, HMC, and GLM-A), the overall algorithm results in significant speedup for this model compared to a conventional single-stage MCMC algorithm because of the fast first-stage evaluation using maximum likelihood estimation and the ability to parallelize large computations over $k$ in the intermediate stage. We refer to this multi-stage method as the GLM exact (GLM-E) method hereafter. 

Both the GLM-A and GLM-E methods result in significant speedup when compared to both the single-stage MCMC algorithm and fully Bayesian multi-stage methods (PG and HMC) due to fast maximum likelihood estimation in the first stage using the \texttt{glm} function in \texttt{R}. Furthermore, the independent samples from the first stage can help prevent sample degeneracy, a common issue that arises in multi-stage MCMC and other Bayesian filtering algorithms \citep{Barreto2025, Scharf2025, Taylor2025}. Figure 1 provides a flow chart that compares the various first-stage sampling strategies and Table 1 compares the strengths of each approach.

\begin{figure}
    \centering
    \includegraphics[width=0.75\linewidth]{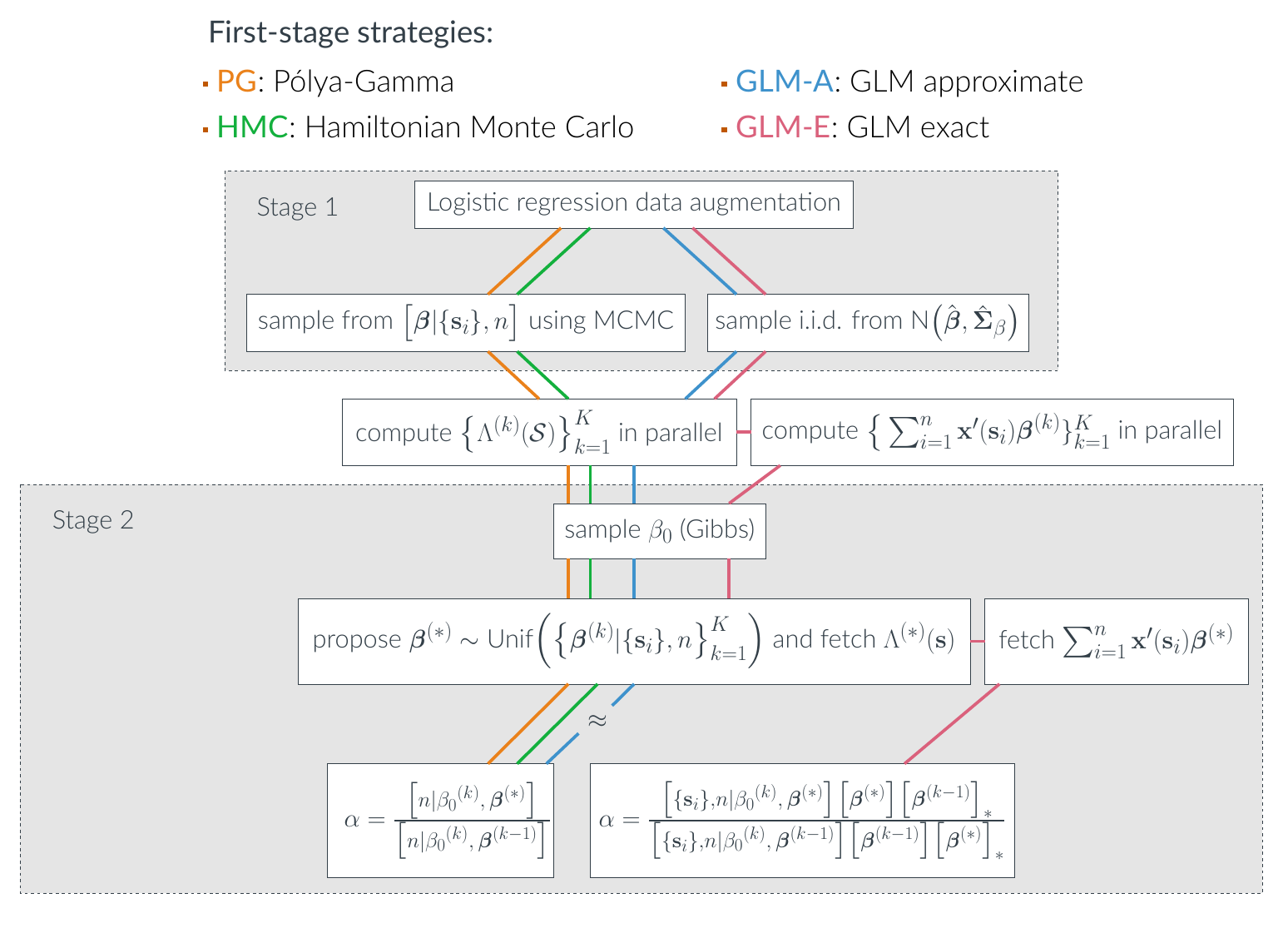}
    \caption{Flowchart of multi-stage algorithm with various first-stage sampling strategies (PG, HMC, GLM-A, GLM-E).}
    \label{fig:placeholder}
\end{figure}

\begin{table}[]
    \centering
    \def\arraystretch{1.5}
    \begin{tabular}{c c c c c}
        \toprule
         Property & PG & HMC & GLM-E & GLM-A \\
         \hline
         Unsupervised & \Checkmark & \Checkmark & \Checkmark & \Checkmark \\
         Fully Bayesian & \Checkmark & \Checkmark & $\times$ & $\times$ \\
         M-H cancellations & \Checkmark & \Checkmark & $\times$ & \Checkmark \\
         Exact Bayesian inference & \Checkmark & \Checkmark & \Checkmark & $\times$ \\
         Independent second stage proposals & $\times$ & $\times$ & \Checkmark & \Checkmark \\
         \bottomrule
    \end{tabular}
    \caption{Comparison of strengths between various first-stage sampling methods.}
\end{table}

\subsection{Compact Window Prediction}

\subsubsection{Posterior Prediction for Total Abundance}

In the compact window setting, recall that we do not observe events that occur outside of the compact observation windows. We denote the number of events outside of the observation windows as $n_0$. If we assume that the point process across the entire spatial domain is an IPP, then by definition, $n_0$ also follows a Poisson distribution:
\begin{equation}
    \big[n_0 | \beta_0, \boldsymbol{\beta}, \{ \mathbf{s}_{ij}\}, \{n_j\}\big] = \text{Pois}\big(\Lambda(\mathcal{S}_0)\big).
\end{equation}
The posterior predictive distribution for $n_0$ is expressed using

\begin{equation}
    \big[ n_0 | \{s_{ij}\}, \{n_j\}\big] = \int\int \big[n_0 | \beta_0, \boldsymbol{\beta}, \{s_{ij}\}, \{n_j\}\big] \big[\beta_0, \boldsymbol{\beta}| \{s_{ij}\}, \{n_j\}\big] \text{d}\beta_0 \text{d}\boldsymbol{\beta}.
\end{equation}

\noindent After obtaining a posterior sample of $\beta_0$ and $\boldsymbol{\beta}$, we obtain a sample from the posterior predictive distribution for $n_0$ (19) via composition sampling by drawing ${n_0}^{(k)} \sim \text{Pois}\big(\Lambda^{(k)}(\mathcal{S}_0)\big)$, where $\Lambda^{(k)}(\mathcal{S}_0)$ is computed using ${\beta_0}^{(k)}$ and $\boldsymbol{\beta}^{(k)}$. Approximating $\Lambda^{(k)}(\mathcal{S}_0)$ still requires numerical quadrature, but can be computed in parallel for $k = 1, \dots, K$. The posterior predictive sample for $n_0$ can then be used to perform a finite population correction for the total number of events in the spatial domain, $\mathcal{S}$: $N^{(k)} = n_0^{(k)} + n$ for $k = 1, \dots, K$, which provides both a point estimate and uncertainty quantification for total abundance.

\subsubsection{Posterior Point Simulation}

We can simulate posterior predictive event locations in the unobserved region $\mathcal{S}_0$ after obtaining estimates for $\lambda(\mathbf{u})$ $\forall \mathbf{u} : \mathbf{u} \in \mathcal{S}_0$ using the Lewis-Shedler method for simulating from an IPP \citep{LewisShedler1979}. Alternatively, a new point realization for the entire study domain $\mathcal{S}$ can be simulated using posterior estimates for $\lambda(\mathbf{u})$ $\forall \mathbf{u} : \mathbf{u} \in \mathcal{S}$. These results can then be used to characterize posterior predictive derived quantities or functions, such as the Ripley's K function or pair correlation function.

\section{Simulation Study}

We demonstrate the multi-stage MCMC approach using a simulation study in a compact window setting. The data were generated from an IPP with an intensity function that is log-linear in terms of three known parameters across the entire study domain (Figure 2). Only the events simulated within the predefined compact windows were used to fit the model with the windowed complete likelihood (3).

\begin{figure}%
    \centering
    \subfloat[\centering]{{\includegraphics[width=0.45\linewidth]{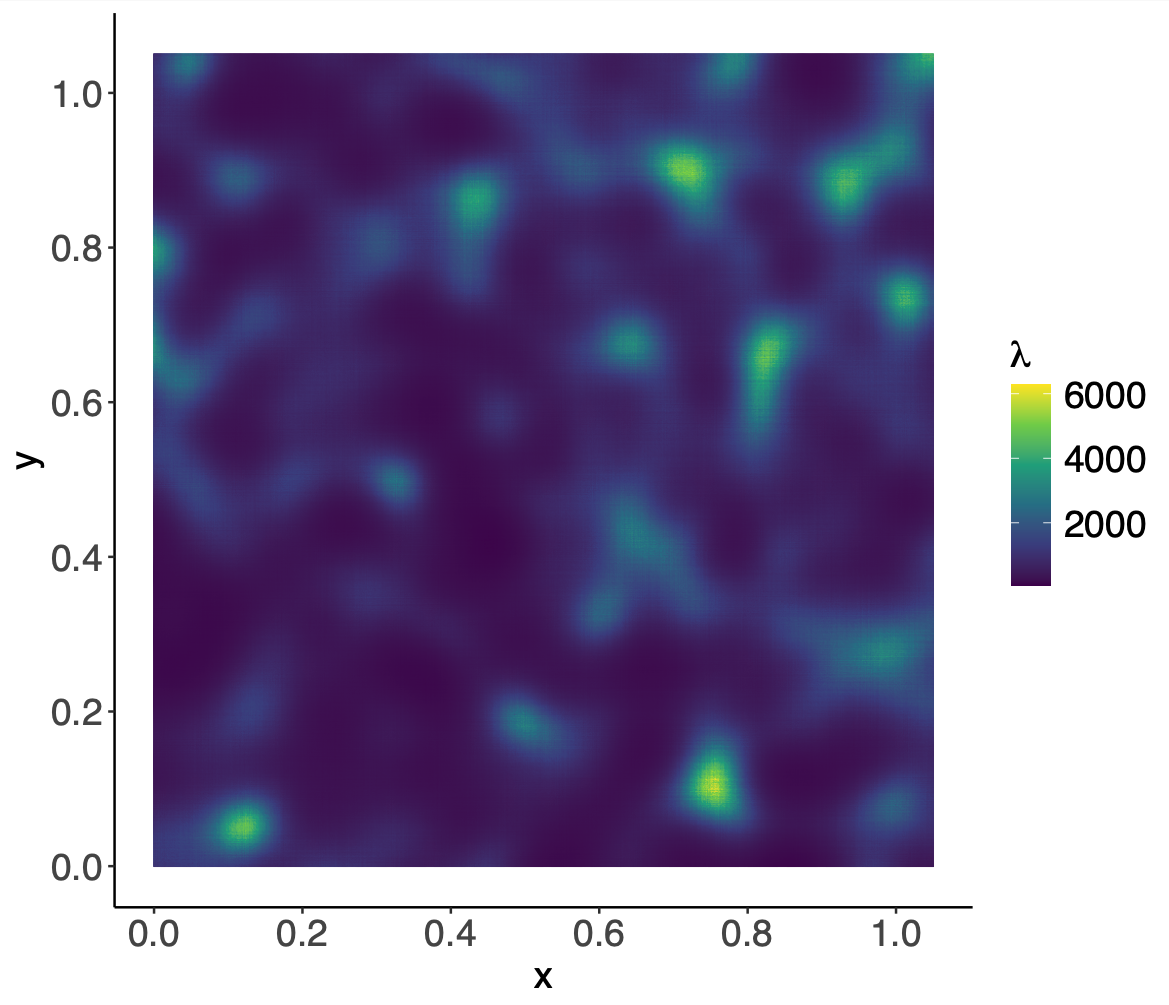} }}%
    \qquad
    \subfloat[\centering]{{\includegraphics[width=0.4\linewidth]{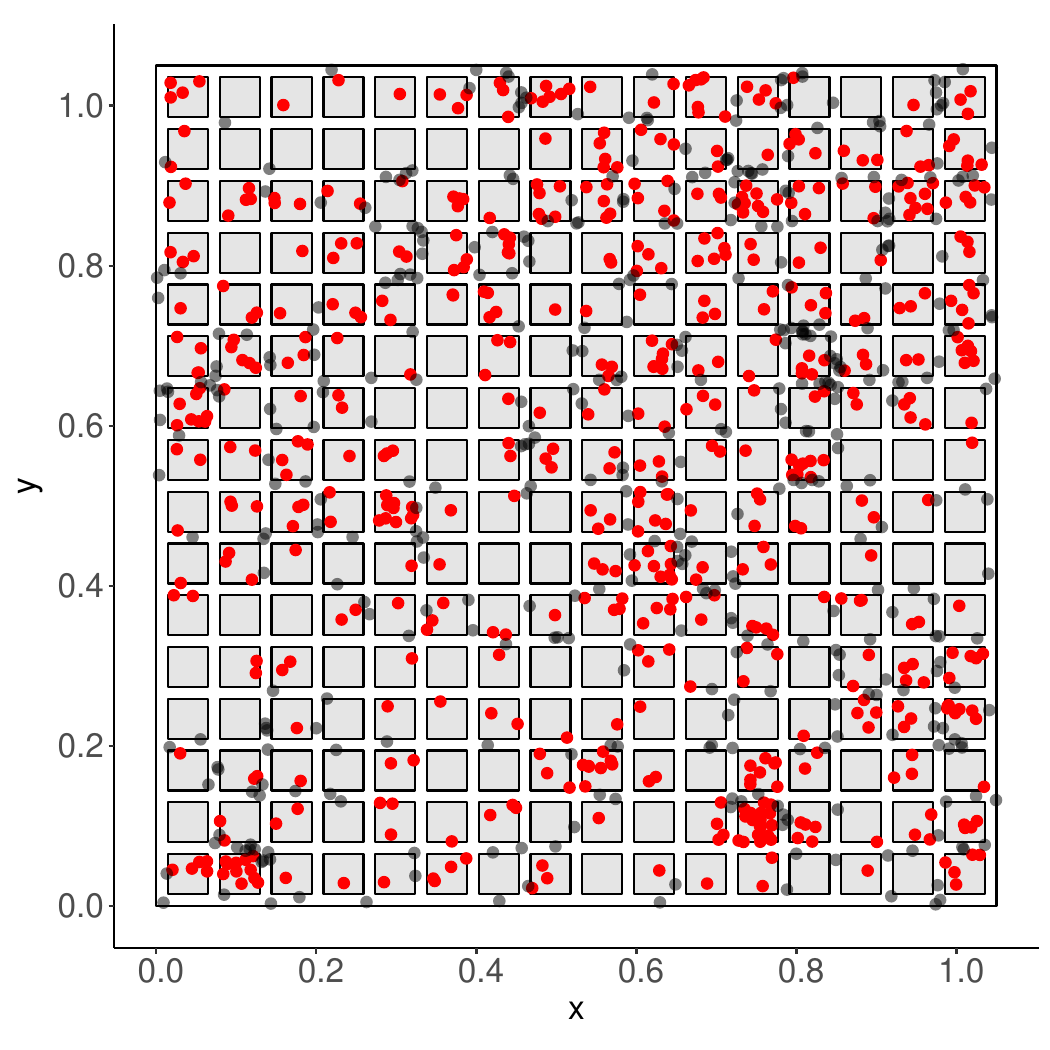} }}%
    \caption{(a) Intensity heat map for simulated data. (b) Realization of simulated points. Points located within the compact windows (red; $n$ = 580) were used to fit the model.}%
\end{figure}

We fit our model to the simulated data using the first-stage sampling strategies described in section 3.2 and a conventional single-stage MCMC algorithm for comparison on a 24-core machine with 3.68 GHz processors and 192 GB of RAM. Figure 3 compares the posterior samples for $\beta_0$ and $\boldsymbol{\beta}$ obtained using the single-stage and the multi-stage MCMC algorithms. It is evident that the marginal posterior distributions for $\beta_0$ and $\boldsymbol{\beta}$ are almost identical among implementation approaches, barring Monte Carlo error and the approximation error from the logistic regression approximation. Similarly, the joint posterior distributions for each pair of model parameters appear nearly indistinguishable between the single-stage method and the GLM-E method (Figure 4).

\begin{figure}
    \centering
    \includegraphics[width=0.8\linewidth]{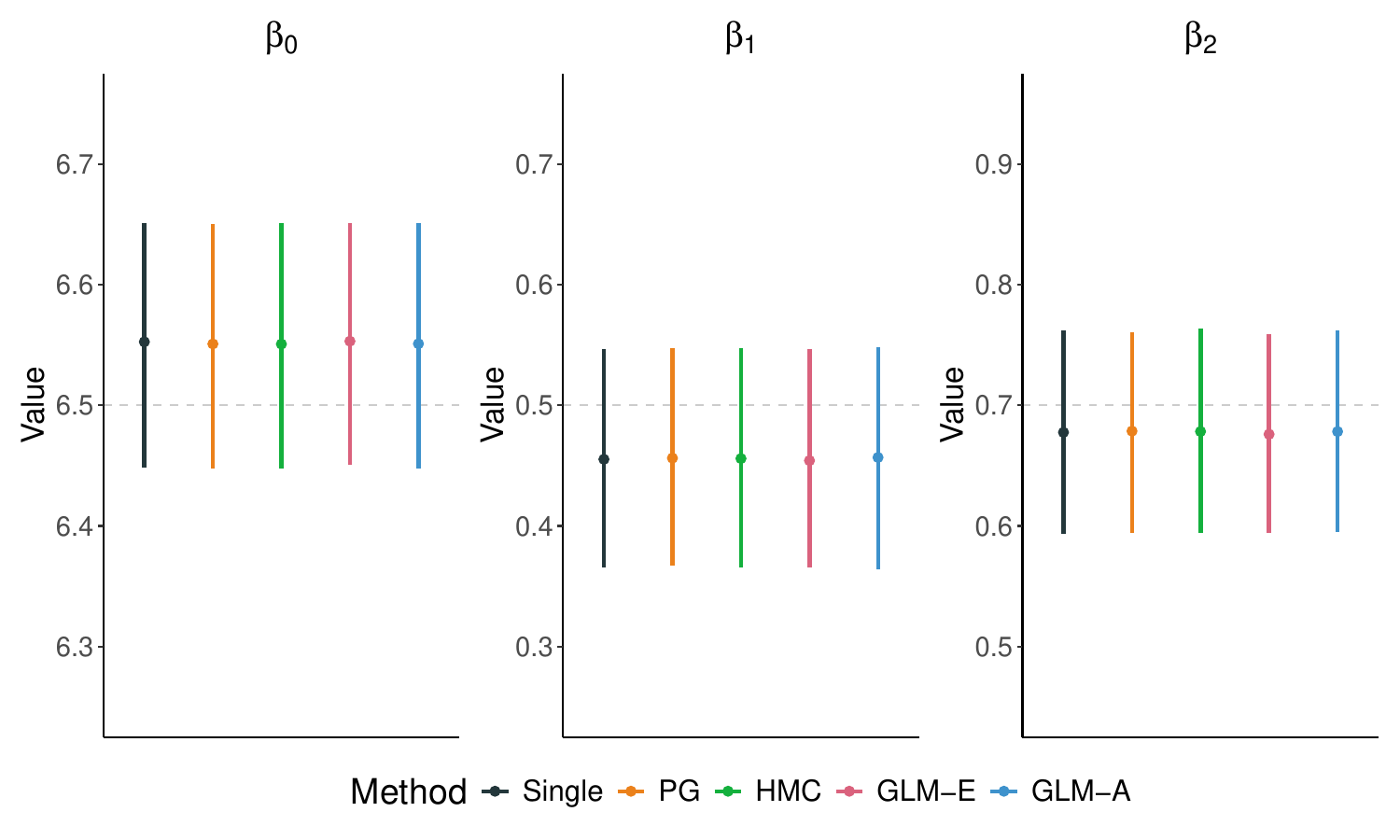}
    \caption{Comparison of marginal posterior 95\% credible intervals for the single-stage method and various multi-stage methods. The true values of the parameters are shown in dashed gray.}
\end{figure}

\begin{figure}
    \centering
    \includegraphics[width=0.8\linewidth]{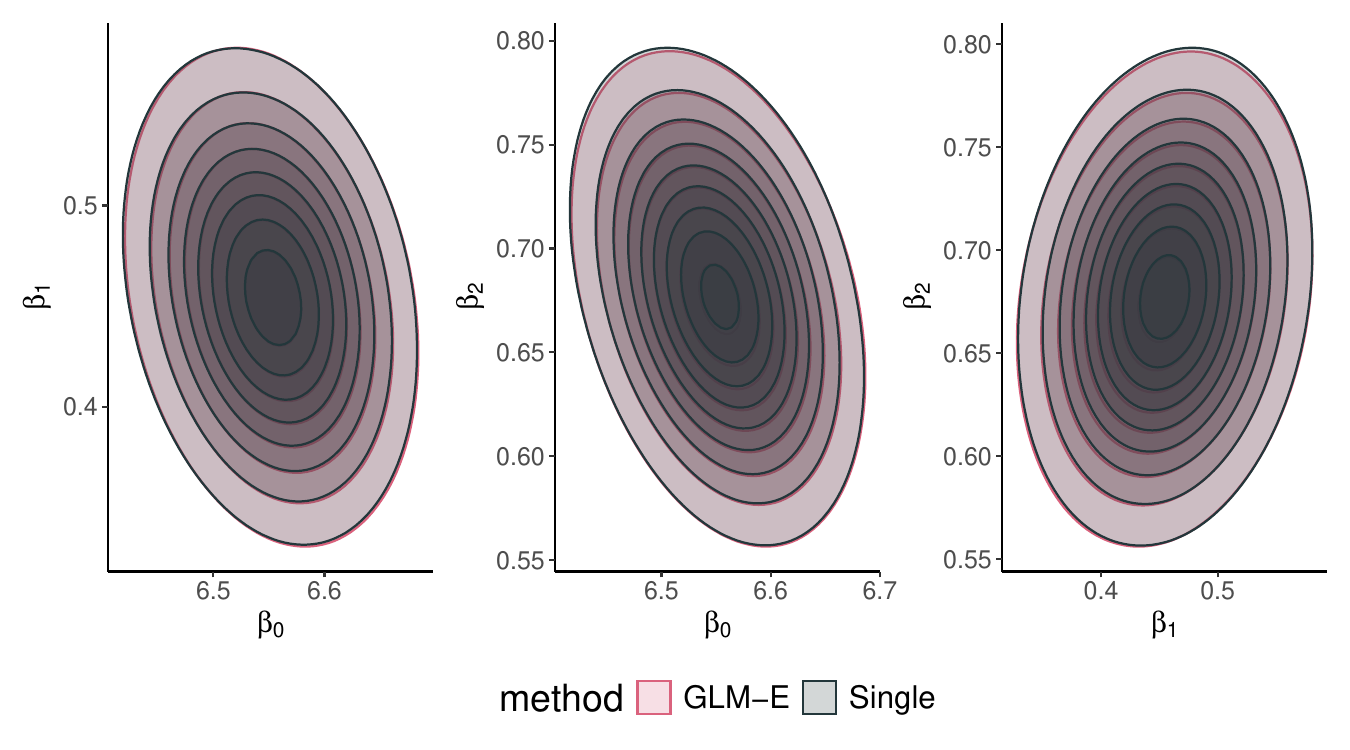}
    \caption{Comparison of bivariate joint posterior distributions for each pair of model parameters. Contours represent kernel density estimates of the joint posterior distributions. The GLM-E contours (pink) are overlaid on top of the single-stage contours (gray).}
\end{figure}

It is worth noting that while the PG, HMC, and GLM-A methods experience some, but often negligible, approximation error from the logistic regression approximation in the first stage, the GLM-E method is unaffected because it does not approximate the conditional likelihood (2) with logistic regression in the M-H correction. Increasing the number of background points $m$ can decrease the logistic regression approximation error, but this incurs significant computational costs for the PG and HMC methods. 

Table 2 compares the seconds per effective sample size between different methods for each parameter. The time recorded for the single-stage method does not include \textit{a priori} parameter tuning, while the remaining multi-stage methods did not require tuning. The Bayesian first-stage methods (PG and HMC) showed an order of magnitude improvement, while the non-Bayesian first-stage methods (GLM-E and GLM-A) showed two orders of magnitude improvement. These results were due to both an improvement in runtime and MCMC effective sample size. All algorithms were fit using $L = 640,000$ quadrature points and the results scale with $L$ and the number of computational cores. Figure 5 shows the posterior predictive distribution for $N$ obtained using the posterior samples from the GLM-E multi-stage method. Similar results were obtained for three additional simulations with varying compact window sizes and latent intensity fields (Appendix B). 

\begin{table}[]
    \centering
    \def\arraystretch{1.2}
    \begin{tabular}{c c c c}
        \toprule
         & $\beta_0$ & $\beta_1$ & $\beta_2$ \\
         \hline
         Single-stage & 0.132 & 0.149 & 0.158 \\
         PG & 0.027 & 0.028 & 0.037 \\
         HMC & 0.014 & 0.014 & 0.018 \\
         GLM-E & 0.001 & 0.001 & 0.001 \\
         GLM-A & 0.001 & 0.001 & 0.001 \\
         \bottomrule
    \end{tabular}
    \caption{Seconds per effective sample for each coefficient and method. Lower values indicate more efficient sampling.}
\end{table}

\begin{figure}
    \centering
    \includegraphics[width=0.5\linewidth]{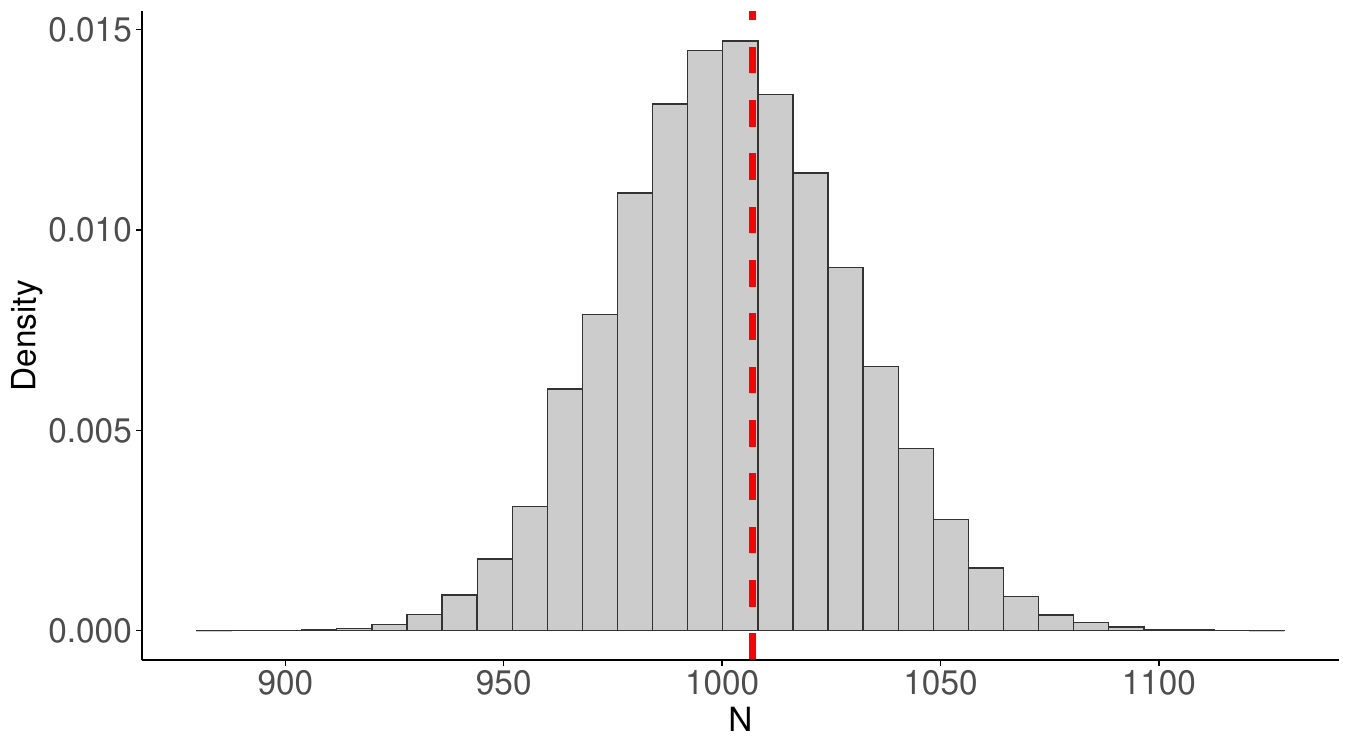}
    \caption{Posterior predictive distribution for $N$ using posterior realizations obtained using the GLM-E method. The true value of $N$ is marked in dashed red.}
\end{figure}

\section{Application: Harbor Seal Pups in Johns Hopkins Inlet}

Johns Hopkins Inlet (Tsalxaan Niyaadé Wool’éex’i Yé; Figure 6) in Glacier Bay National Park is one of several tidewater glacial fjords in southeastern and south central Alaska that collectively host some of the largest seasonal aggregations of harbor seals (\textit{Phoca vitulina}) in the world \citep{Jansen2015, Womble2020}. Harbor seals use the icebergs in these tidewater glacial fjords as haul out sites for resting, pupping, molting, and avoiding predators \citep{Womble2020}. Harbor seal abundance and space use in Johns Hopkins Inlet are of particular interest to management at Glacier Bay National Park because they help inform policy that aims to minimize potential seal disturbance while also allowing park visitors access to tidewater glaciers, one of the founding mandates of the park \citep{NPS2010, NPS2025}.

Harbor seal locations in Johns Hopkins Inlet were observed using aerial imagery surveys in June and August, the height of pupping and molting, respectively \citep{Womble2020, Womble2021}. Non-overlapping photos were taken along flight transects to avoid double-counting seals and provide systematic sampling of the entire inlet. Each image covered approximately 80 m $\times$ 120 m of surface area. We refer to the corresponding geo-referenced region as an image footprint hereafter. After imagery data were collected, a trained observer reviewed each digital image and annotated the spatial locations of adult and pup seals hauled out on ice. Further details on the survey methods and data processing are presented in \citet{Womble2020}.

Previous studies analyzing these data have focused on modeling harbor seal abundance for the entire inlet or using coarse spatial grids. For example, \citet{Womble2020} incorporated the aerial imagery survey data along with other sources of data to obtain estimates and uncertainty quantification for harbor seal abundance. \citet{Womble2021} fit a multivariate conditional autoregressive (MCAR) model to summarize spatially explicit abundance on a 200 m $\times$ 300 m grid ($L = 434$ total grid cells). \citet{Jansen2015} used harbor seal data from a similar aerial survey to fit a generalized additive model (GAM) and summarized spatially explicit abundance on a grid of 400 m $\times$ 400 m cells ($L = 779$ total grid cells). Previous studies on spatially explicit abundance of harbor seals in Johns Hopkins Inlet have not explored modeling the locations as a spatial point process directly. Modeling these data as an SPP provides unique advantages because it allows for informed simulation of potential seal locations outside the image footprints using an established statistical framework. Furthermore, the multi-stage algorithm makes fitting the model on a fine spatial grid more computationally flexible and feasible, especially with the availability of distributed computing resources.

We fit a Bayesian SPP model to harbor seal pup haul out location data from 21 June 2007 ($n = 452$; Figure 7) using the aforementioned multi-stage MCMC approaches. We modeled harbor seal pups because they are the most vulnerable segment of the population and are especially susceptible to disturbance from vessels \citep{Mathews2016}. Furthermore, pups are hauled out on ice and observable with high probability at the time the survey was conducted (i.e., in the month of June and in the early afternoon) based on secondary analysis of wet-dry tag data of harbor seal pups in Disenchantment Bay, Alaska. Our study domain spanned from the Johns Hopkins Glacier terminus to Jaw Point, which is approximately 9km long and 1.5-2km wide (Figure 6a). The image footprints for this survey cover approximately half of the study domain. We assume perfect detection of hauled out pups within the image footprints and no detection outside the image footprints, giving rise to compact observation windows. Additionally, due to the systematic surveying within the study domain, we assume the unobserved areas in the inlet to be ignorable missing data (i.e., the covariate values and relationship between covariates and pup locations do not differ significantly between observed and unobserved areas). Using the pup locations as point events, we fit the SPP model using the windowed complete likelihood (3).

\begin{figure}
    \centering
    \includegraphics[width=0.7\linewidth]{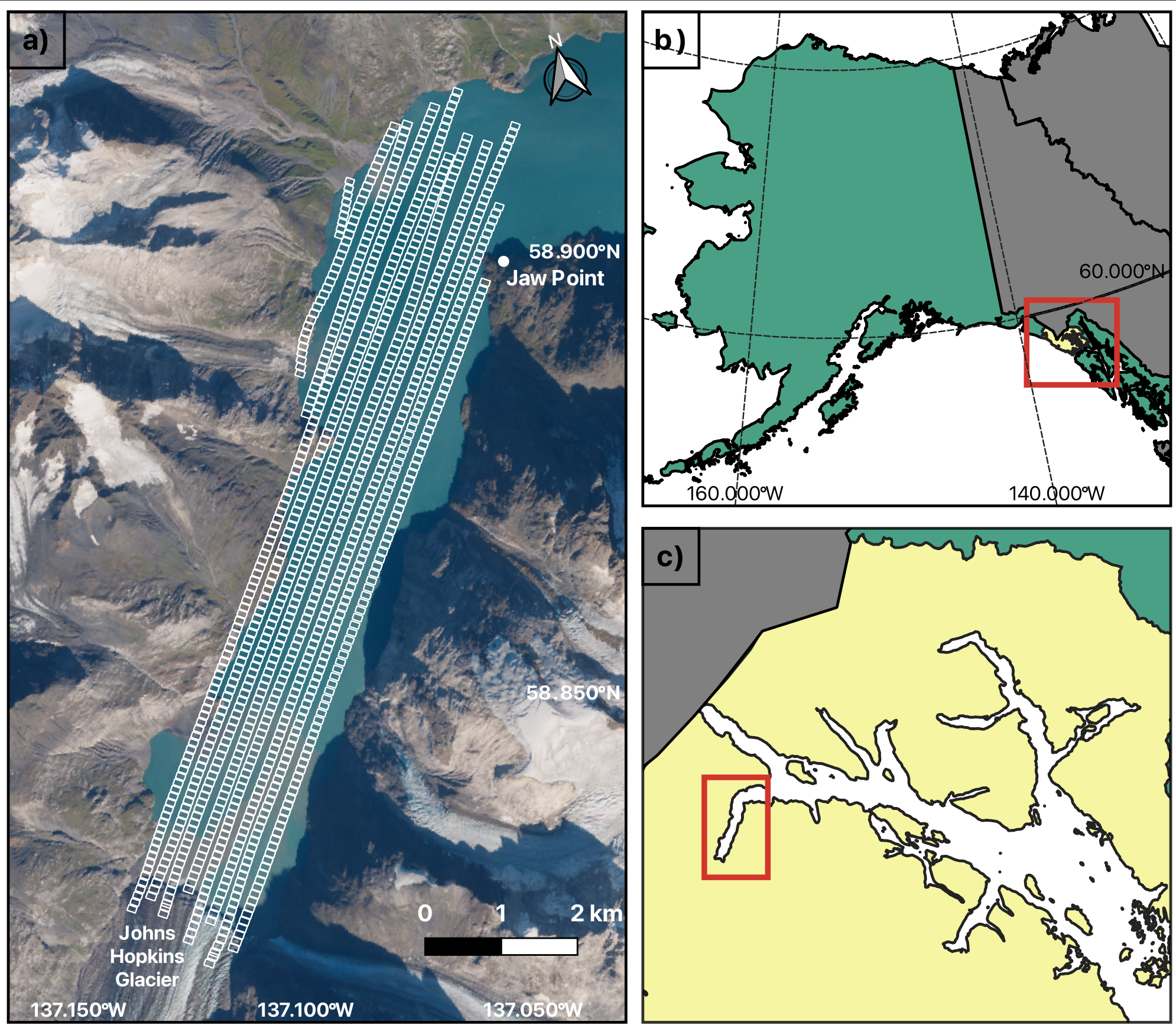}
    \caption{(A) A Sentinel-2 satellite image of Johns Hopkins Inlet in 2018. Each white rectangle represents an image footprint from the 21 June 2007 survey. (B) and (C) show the location of Glacier Bay National Park and Johns Hopkins Inlet, respectively, in red.}
\end{figure}

\begin{figure}
    \centering
    \includegraphics[width=0.5\linewidth]{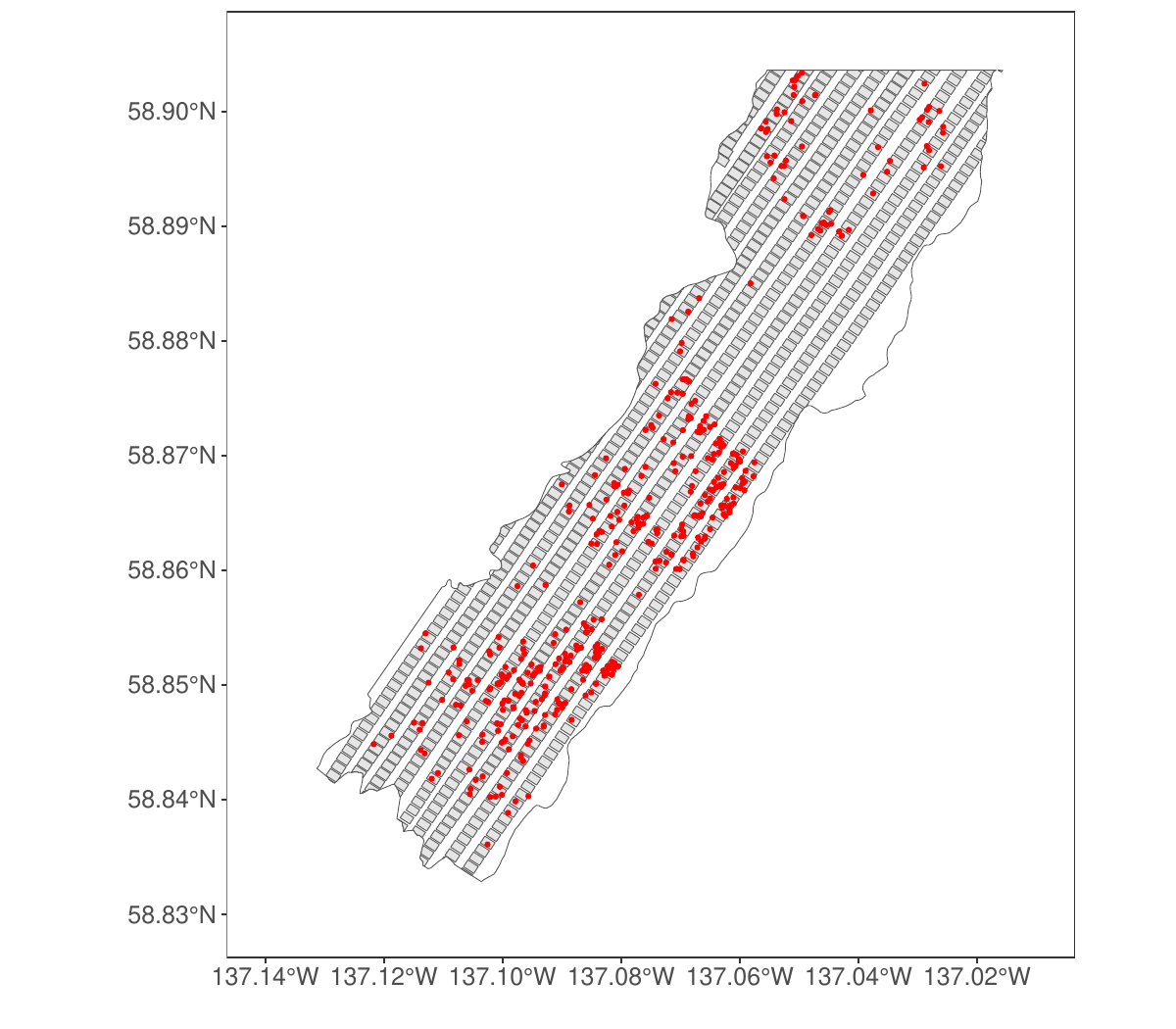}
    \caption{Observed harbor seal pup haul out locations (red) on 21 June 2007 ($n = 452$). The large polygon represents the survey domain and each gray rectangle represents an image footprint.}
\end{figure}

We fit the model using three covariates: ice proportion, bathymetry, and distance to glacier terminus (Appendix C). The ice data were represented as iceberg outlines within the image footprints and summarized as ice proportions to allow for spatial prediction outside the image footprints (Appendix D). Each covariate was summarized on the same raster grid with a spatial resolution of approximately 10 m $\times$ 10 m ($L = 304,945$ total grid cells). The grid centers were used as quadrature points throughout the analysis.

\subsection{Accounting for Additional Spatial Heterogeneity}

From preliminary analysis and model checking, we found that using a log-linear combination of the three covariates to model the IPP intensity was insufficient in capturing the heterogeneity of the observed harbor seal pup data (Appendix E). We first discuss how we modified the model to account for additional spatial heterogeneity for the harbor seal pup model in addition to alternative strategies and their compatibility with the multi-stage algorithm.

\subsubsection{Neural Network Basis Expansion}

To increase model flexibility and allow for more complex relationships between covariates and intensity, we constructed a basis representation of $\mathbf{X}$, the $p$ $\times$ $L$ design matrix consisting of the covariate values at each grid center. Such practice is common in spatial modeling, with the popularity of fixed-rank \citep{Cressie2008} and radial \citep{Buhmann2000} basis functions. Of the basis functions typically used to account for spatial heterogeneity \citep{Hefley2017, WikleZammit-Mangion2023}, we found that a single-hidden layer feedforward neural network (SLFNN) was able to capture the relationship between the covariates and pup space use. Additionally, we found that the Extreme Learning Machine (ELM) scheme \citep{Huang2006} was sufficient for constructing the neural network (NN) weights and resulted in efficient computation in the multi-stage algorithm. 

To incorporate the NN basis expansion in the multi-stage algorithm, we first trained the NN in the first stage when fitting the logistic regression approximation. Using our samples generated from the Berman-Turner device $\Big\{\big(\mathbf{x}(\mathbf{s}_i), y(\mathbf{s}_i)\big)\Big\}_{i = 1}^{\tilde{n}}$, we modeled the relationship between $\mathbf{x}(\mathbf{s}_i)$ and $\tilde{y}_i = \text{logit}(p_i)$ as a linear relationship using $q$ hidden nodes
\begin{equation}
   \sum_{j = 1}^q \tilde{\beta}_j g(\boldsymbol{\omega}_j \cdot \mathbf{x}_i) := \tilde{y}_i,
\end{equation}
where $g(\cdot)$ represents a user-specified activation function, $\boldsymbol{\omega}_j$ is a $p \times 1$ vector of weights, and $\tilde{\beta}_j$ represents the coefficient associated with the $j$th hidden node. The quantity in (20) can be summarized compactly as 
\begin{equation}
    \mathbf{W}\tilde{\boldsymbol{\beta}} = \tilde{\mathbf{y}}_i,
\end{equation}
where
\begin{equation}
    \mathbf{W} = \begin{bmatrix}
        g(\boldsymbol{\omega}_1 \cdot \mathbf{x}_1) & \cdots & g(\boldsymbol{\omega}_q \cdot \mathbf{x}_1)\\
        \vdots & \cdots & \vdots \\ 
        g(\boldsymbol{\omega}_1 \cdot \mathbf{x}_{\tilde{n}} ) & \cdots & g(\boldsymbol{\omega}_q \cdot \mathbf{x}_{\tilde{n}} )
    \end{bmatrix}_{\tilde{n} \times q}.
\end{equation}
The matrix $\mathbf{W}$ is referred to as the hidden layer output matrix. Instead of learning the weights, $\boldsymbol{\omega}_j$ for $j = 1, \dots, q$ using gradient-based methods, the ELM scheme samples random weights from a user-specified distribution for efficient computation \citep{Huang2006}. After we computed $\mathbf{W}$ using the randomly generated weights, we used it as a basis expansion for $\mathbf{X}$ (i.e., each hidden node represents a basis vector), which then served as a new design matrix for the first-stage logistic regression problem. Aside from replacing $\mathbf{X}$ with $\mathbf{W}$, the intermediate and second stages in the multi-stage algorithm remain unchanged.

For the 21 June 2007 harbor seal pup data, we used the Gaussian Error Linear Units (GELU) activation function \citep{hendrycks2016gelu} defined as $g(\mathbf{x}) = \mathbf{x}\Phi(\mathbf{x})$, where $\Phi(x)$ is a standard Gaussian cumulative distribution function, and generated 100 hidden layer output matrices for $q = 5$ and weights sampled from a standard Gaussian distribution. We selected the $\mathbf{W}$ matrix with the optimal AIC based on (20) for the basis expansion. We then used the GLM-E multi-stage algorithm to obtain $K = 100,000$ posterior realizations for $\tilde{\boldsymbol{\beta}}$ on the same machine as the simulation study (24-core, 3.68 GHz processors and 192 GB of RAM), which required approximately three minutes for the entire multi-stage algorithm. The posterior realizations were used to simulate points across the entire study domain using the method described in section 3.3.2. The simulated points were summarized in a grid of posterior mean abundance (Figure 8). We validated our model using posterior predictive model-checking and did not observe large discrepancies between the simulated and observed data (Appendix E). 

\begin{figure}
    \centering
    \includegraphics[width=0.6\linewidth]{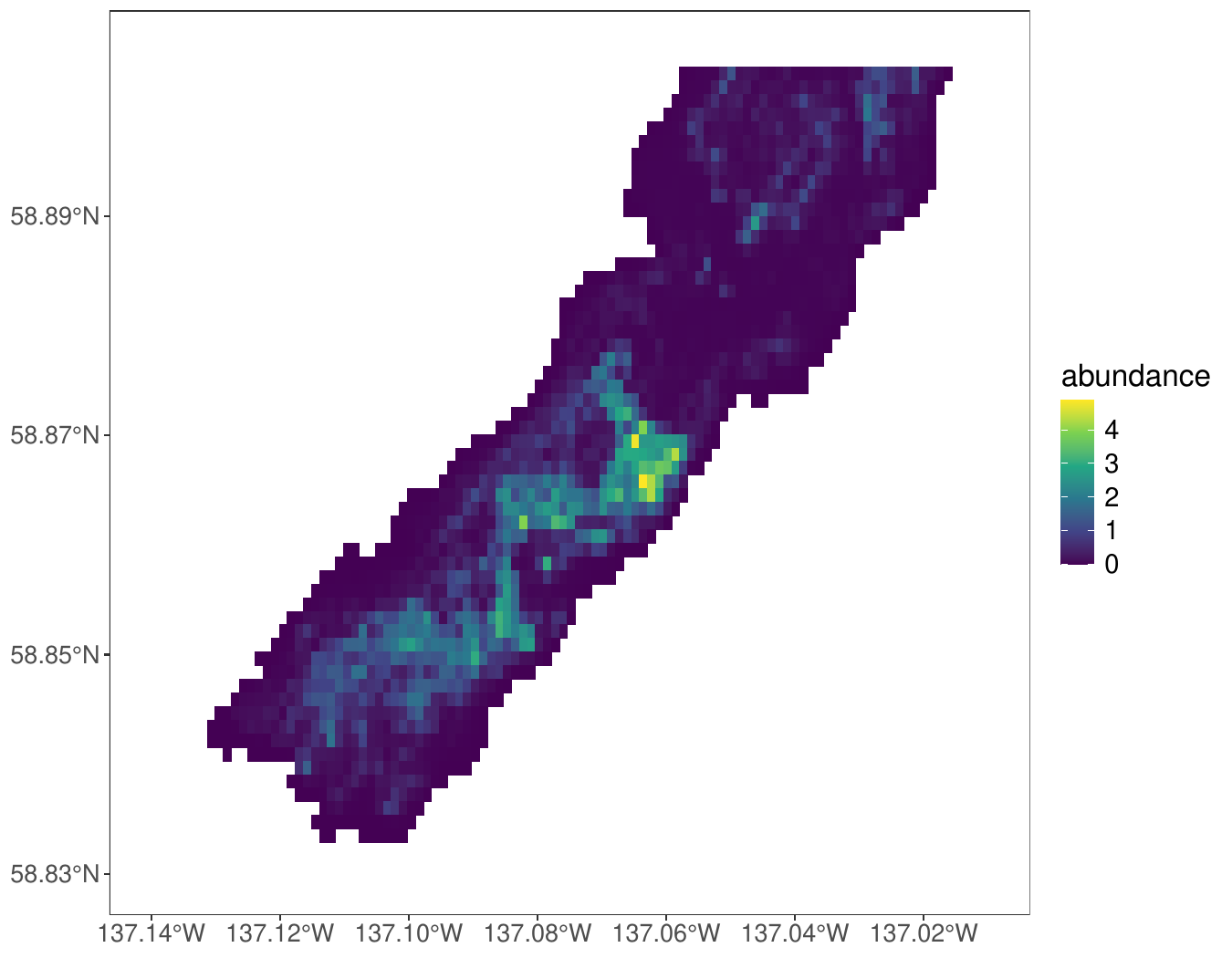}
    \caption{Spatially explicit posterior mean abundance of simulated harbor seal pup haul out locations on 21 June 2007.}
\end{figure}

\begin{figure}
    \centering
    \includegraphics[width=0.5\linewidth]{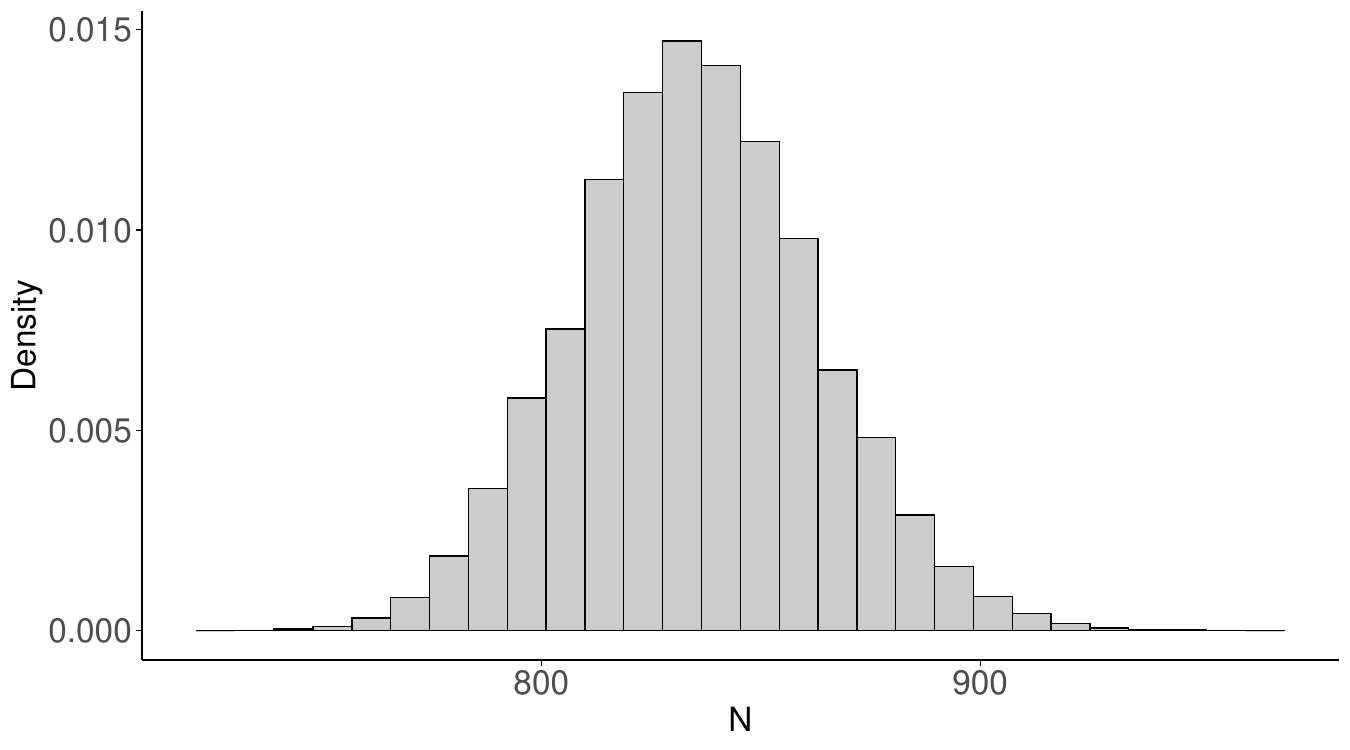}
    \caption{Posterior predictive distribution for total abundance of hauled out seal pups $N$ obtained using the GLM-E posterior realizations.}
\end{figure}

When compared to the observed point pattern in Figure 7, Figure 8 indicates that greater abundance values were simulated in areas of high observed pup concentrations in addition to surrounding areas that were originally unobserved. Areas of greater simulated abundance values are also consistent with areas of higher ice proportions and more negative bathymetry values (i.e., areas of deeper water), supporting previous studies on harbor seal behavior in Southeast Alaska \citep{Montgomery2007, Kaluzienski2023}. Additionally, although we did not include this as a covariate, greater abundance values were simulated in areas where eddies typically form in the inlet, which is consistent with findings that harbor seals prefer relatively slow-moving icebergs during the pupping season \citep{Kaluzienski2025}. The posterior predictive distribution for total abundance $N$ was generated using the procedure in 3.3.1 (Figure 9). The posterior mean and standard deviation of $N$ were 835.2 and 27.7, respectively.

\subsubsection{Alternative Strategies for Increased Model Flexibility}

Although we found the NN basis expansion approach to be both sufficient for capturing complex spatial patterns, such as the harbor seal pup data, and computationally efficient within the multi-stage computing algorithm, alternative strategies for increased model flexibility exist and may be applicable. For instance, fixed-rank or radial basis functions may be used instead of the NN basis functions. \citet{Dovers2024} showed that a similar basis function approach can be used to approximate an LGCP model and only requires fitting a generalized linear mixed model (GLMM) in \texttt{R}. Fitting the multi-stage algorithm using an SPP model with a true latent Gaussian process to capture additional spatial heterogeneity as in an LGCP model \citep{Brix2001} is achievable, but introduces additional computational complexities associated with learning parameters for a high-dimensional latent field, which is the subject of ongoing research.

\section{Discussion}

Recent innovation in technology and survey techniques have led to significant improvements in the availability and spatial resolution of datasets, making the development of computationally-efficient and scalable SPP models especially crucial. Many useful methods have been developed to facilitate flexible and efficient SPP modeling \citep{Baddeley2010, WartonShepherd2010, Aarts2012, FithianHastie2013} for conditional IPP likelihoods as in (2). Furthermore, developments in using the Laplace approximation to approximate Bayesian SPP model marginal distributions have led to significant computational improvements \citep{Rue2009, Illian2012} for fitting LGCP models. However, fitting an SPP model that both uses the complete likelihood (1), which accounts for the stochasticity associated with $n$, and provides exact Bayesian inference for the joint posterior distribution has remained computationally challenging.

More recently, recursive Bayes methods for fitting ecological statistical models have gained traction and have been shown to substantially improve computational efficiency \citep{Hooten2021, Hooten2023, Hooten2024, Leach2022, McCaslin2020}. We showed that the complete likelihood for an IPP model (1) can be similarly fit in stages where the coefficients are first estimated using the likelihood conditioned on $n$ (2), then corrected with the Poisson likelihood for $n$ (8). Although sample degeneracy is a commonly cited issue for recursive Bayes and other sequential filtering algorithms \citep{Barreto2025, Scharf2025, Taylor2025}, our multi-stage algorithm did not suffer from this phenomenon due to only having two stages and the ability to sample the additional learned parameter in the second stage, $\beta_0$, using a Gibbs sampler. This is especially true for the non-Bayesian first stage strategies, GLM-A and GLM-E, because the transient posterior realizations from (16) are independent. In practice, the acceptance rate for $\boldsymbol{\beta}$ in the second stage ranged from 0.25 to 0.7 in our simulations and case study.

Fitting the IPP model in stages provides many benefits including increased flexibility in the first stage. We showed that fitting the first stage is approximately equivalent to fitting a model using the conditional likelihood (2), which is well approximated using logistic regression. This approximation provides various ways to fit the first stage, including user-friendly GLM \texttt{R} packages, making fitting Bayesian SPP models more accessible to a broader audience. Additionally, the multi-stage methods we presented do not require algorithm supervision and result in improved mixing for MCMC chains when compared to a single-stage MCMC algorithm, especially for the non-Bayesian first-stage approaches, GLM-E and GLM-A. Furthermore, all aforementioned multi-stage methods leverage parallel computing resources to pre-compute computationally intensive values, allowing the method to scale with the number of computer cores and increase the feasibility of fitting an IPP model using a large number of quadrature points.

The multi-stage construction of the algorithm also allows practitioners to take a meta-analytic approach to adjust model output that was originally fit using the conditional IPP likelihood (2) by incorporating data that are already available. The original MCMC sample (or approximate Gaussian distribution if non-Bayesian) can be treated as the transient posterior realizations in the multi-stage algorithm. Proceeding with the intermediate and second stages adjusts the estimates to account for the stochasticity associated with $n$ and provides learning about the $\beta_0$ parameter, which was non-identifiable during the original model fitting. The resulting model output provides exact Bayesian inference using the complete likelihood in (1).

Restricting the IPP intensity function to be log-linear in the available covariates may not be sufficient to capture spatial heterogeneity in complex spatial patterns, such as the harbor seal pup data in Figure 6. In section 5.1, we outlined ways to account for additional spatial heterogeneity in the model, including using a NN basis expansion, for which training is compatible and efficient within the multi-stage algorithm. Furthermore, adding x- and y-coordinates to the basis expansion functions similarly to the LGCP in that they account for spatial heterogeneity that is unexplained by the observed covariates \citet{Dovers2024}. One main limitation of modeling the points as an IPP or LGCP is that points are assumed conditionally independent given the spatial covariates (and latent field for the LGCP). Basis functions and/or latent fields can allow for flexible estimation of complex intensity fields, but additional dependence among points may be necessary to reasonably capture extreme forms of spatial heterogeneity such as clustering due to social structure. Exploratory data analysis on the observed point pattern and model validation techniques concerned with the second-order properties of a point pattern \citep{Illian2008} should be explored during analysis. Future directions of this work include extending the multi-stage procedure for self-exciting point processes such as the Hawkes process \citep{Hawkes1971}. 

Finally, we showed how this method can be extended to the compact window setting where a posterior predictive distribution for the total abundance $N$ can be obtained. Furthermore, this method allows for intensity estimation and point simulation that lie outside the observation windows, which can be useful for downstream analyses. 

\section{Acknowledgments}
This research was funded by the National Science Foundation Graduate Research Fellowship Program and the National Park Service. The authors thank Paul Conn, Brett McClintock, Justin Van Ee, Daniel Barreto, and Myungsoo Yoo for helpful discussions and insights. We also thank Jason Amundson and John Jansen for providing relevant data.

\newpage

\bibliographystyle{cas-model2-names}
\bibliography{spatial_refs}

\newpage

\section*{Appendix A}

Recall that the complete likelihood is given by 
$$ \big[\{ \mathbf{s}_{i}\}, n | \beta_0,  \boldsymbol{\beta}\big] = \frac{\prod_{i = 1}^{n} \lambda(\mathbf{s}_{i})}{n! \exp\bigl( \Lambda(\mathcal{S}) \bigr)}.$$
We can rewrite the complete likelihood using $\zeta := e^{\beta_0}$:
\begin{align*}
    \big[\{ \mathbf{s}_{i}\}, n | \beta_0,  \boldsymbol{\beta}\big] &= \frac{\prod_{i=1}^n \zeta \exp\{ \mathbf{x}^{\prime}(\mathbf{s})\boldsymbol{\beta}\} }{n! \exp \big\{ \zeta \Lambda(\mathcal{S}) \big\} }\\
    &= \frac{1}{n!} \zeta^n \exp \Big\{ -\zeta \Lambda(\mathcal{S}) + \exp\{ \mathbf{x}^\prime(\mathbf{s})\boldsymbol{\beta} \} \Big\}.
\end{align*}
Then, using a vague $\text{Gamma}(a,b)$ prior with shape-rate parameterization, we can obtain a closed-form full-conditional distribution for $\zeta$:
\begin{align*}
    \big[ \zeta | \{ \mathbf{s}_i \}, n, \boldsymbol{\beta}\big] &\propto [\{ \mathbf{s}_i \}, n| \zeta, \boldsymbol{\beta}] [\zeta]\\
    &\propto \zeta^n \exp \big\{ -\zeta \Lambda(\mathcal{S}) \big\} \zeta^{a-1}\exp\{-b\zeta\}\\
    &\propto \text{Gamma}\big(a + n, b + \Lambda(\mathcal{S})\big).
\end{align*}

Recall that $\Lambda(\mathcal{S})$ in the full-conditional update can be pre-computed in the previous stage in the multi-stage MCMC algorithm. Thus, sampling from the full-conditional distribution for $\zeta$ in the second stage is fast. 

\section*{Appendix B}

We conducted simulations using two distinct compact windowed constructions in which the windows cover approximately half of the study domain area. For each of the windowed constructions, we fit the IPP model with the windowed complete likelihood (3) using a single-stage MCMC algorithm and each multi-stage method to data generated from two distinct latent intensity fields. The simulation study with small windows and point pattern 1 is presented in section 4, while the remaining three simulations are shown below.

\subsection*{Small windows, point pattern 2}

\begin{figure}[H]
    \centering
    \subfloat[\centering]{{\includegraphics[width=0.45\linewidth]{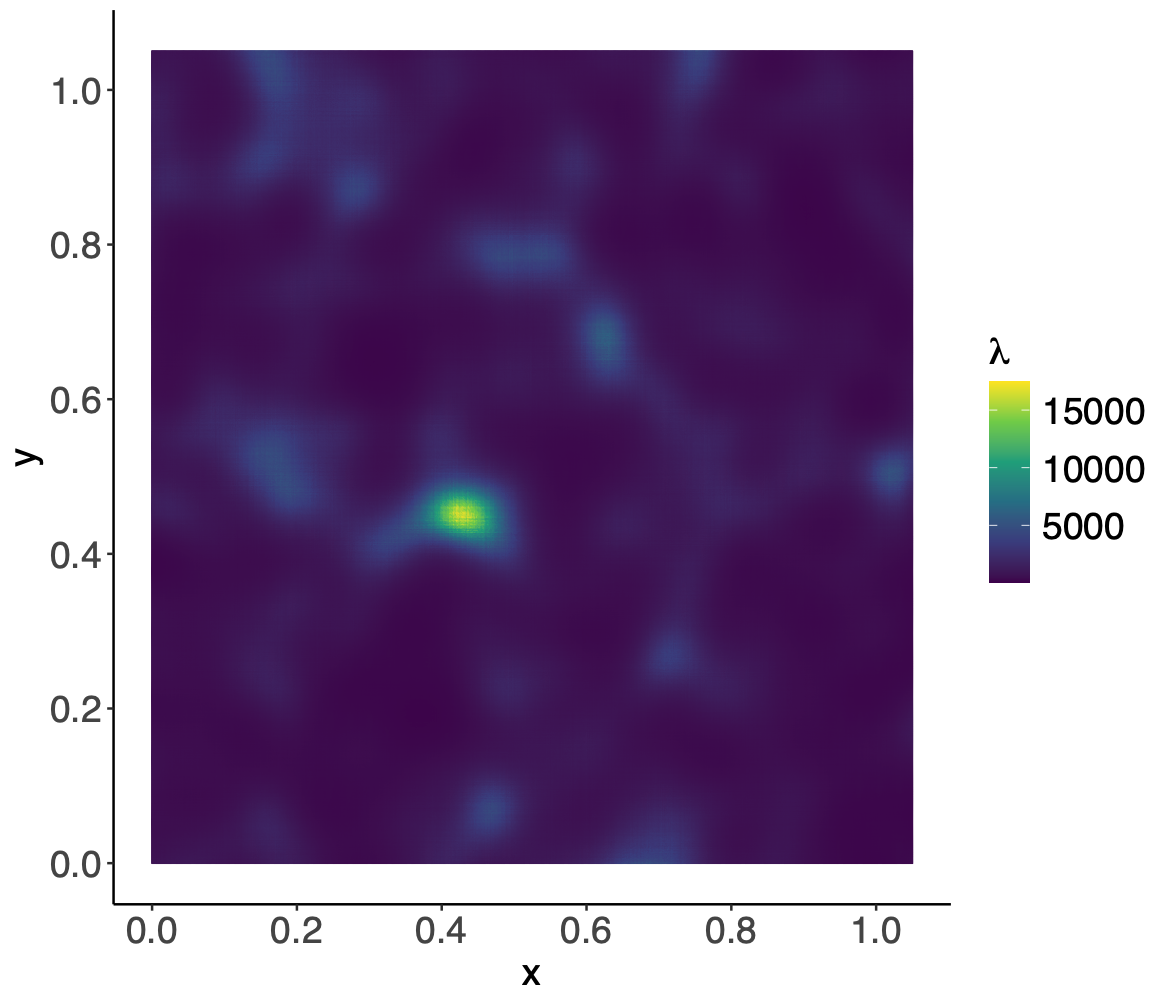} }}%
    \qquad
    \subfloat[\centering]{{\includegraphics[width=0.4\linewidth]{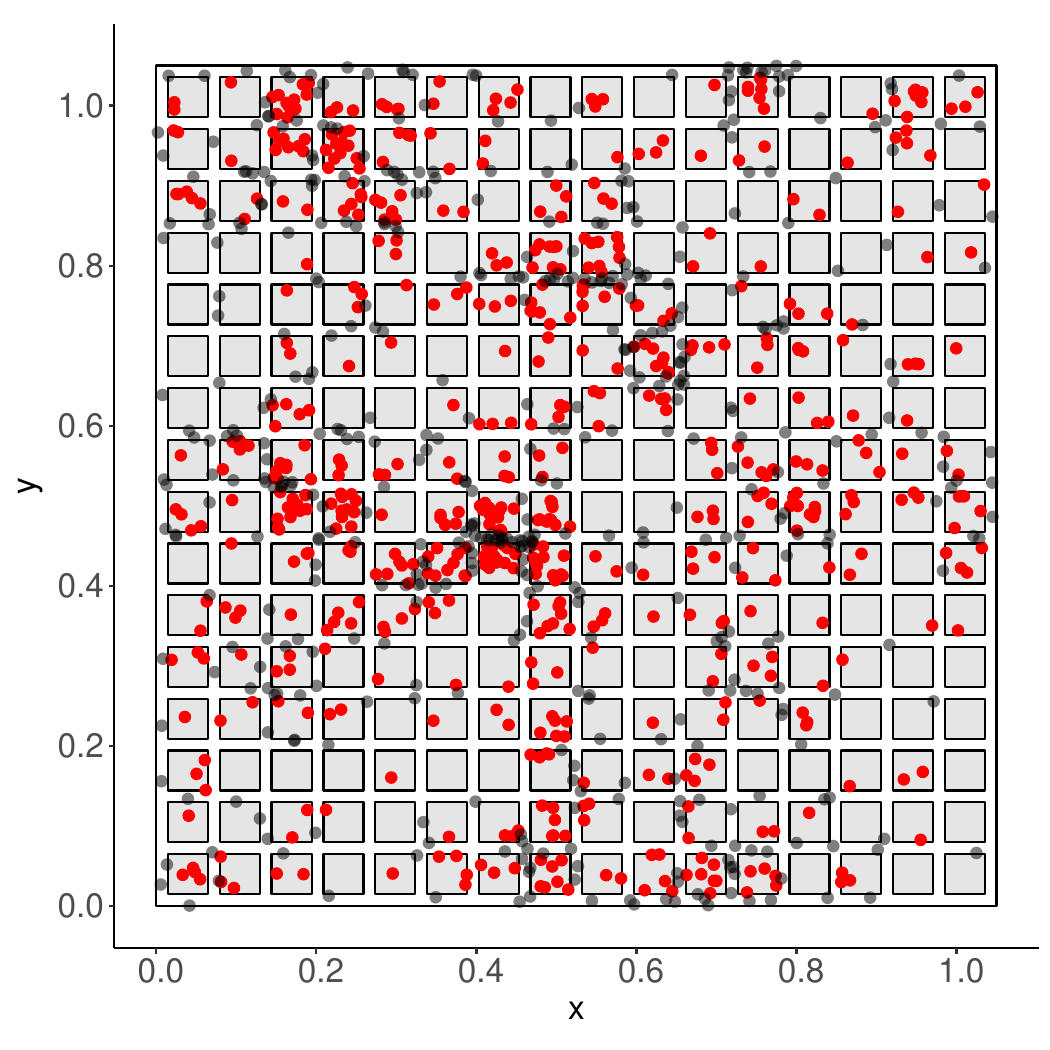} }}%
    \caption{(a) Intensity heat map for simulated data. (b) Realization of simulated points. Points located within the compact windows (red; $n$ = 617) were used to fit the model.}%
\end{figure}

\begin{figure}[H]
    \centering
    \includegraphics[width=0.8\linewidth]{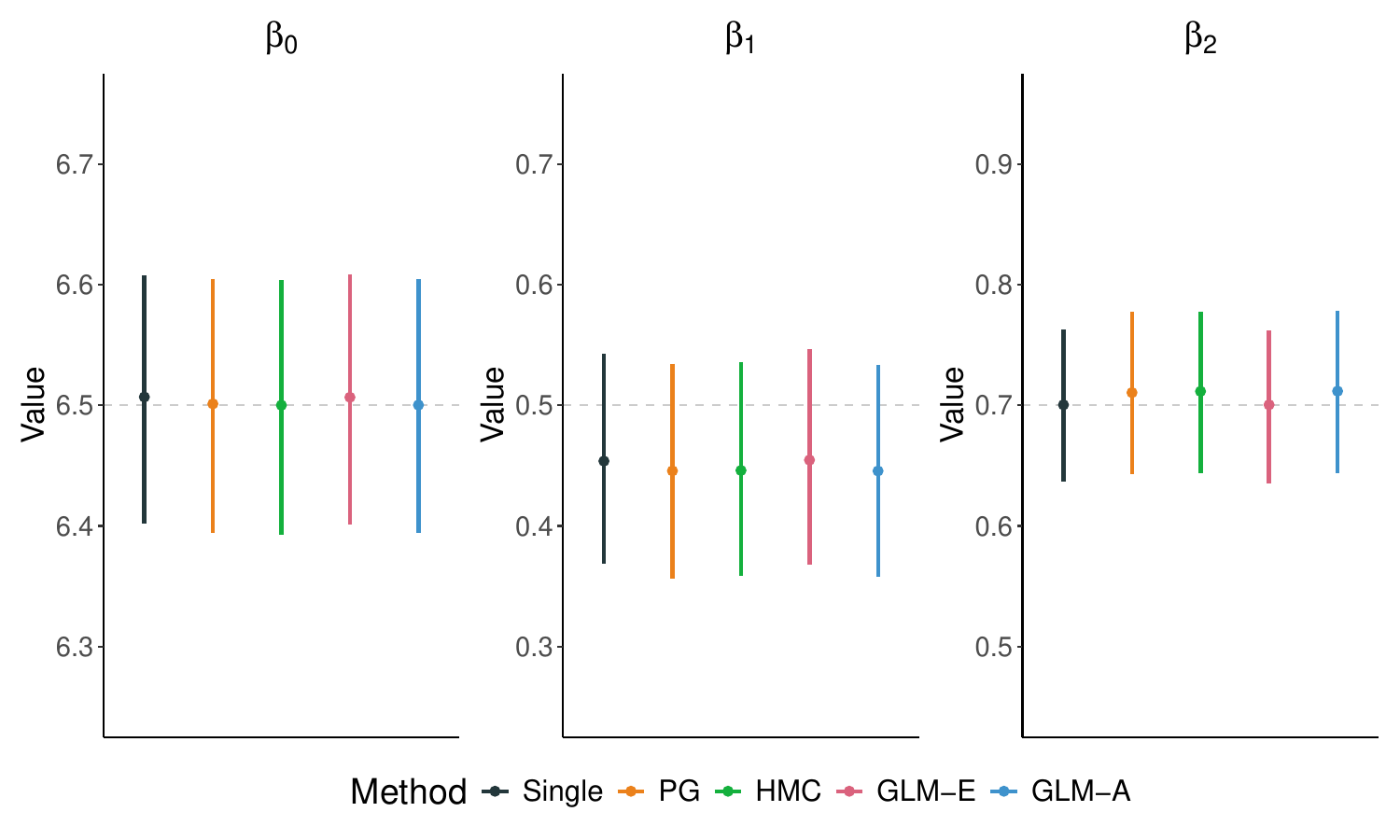}
    \caption{Comparison of marginal posterior 95\% credible intervals for the single-stage method and various multi-stage methods. The true values of the parameters are shown in dashed gray.}
\end{figure}

\begin{figure}[H]
    \centering
    \includegraphics[width=0.8\linewidth]{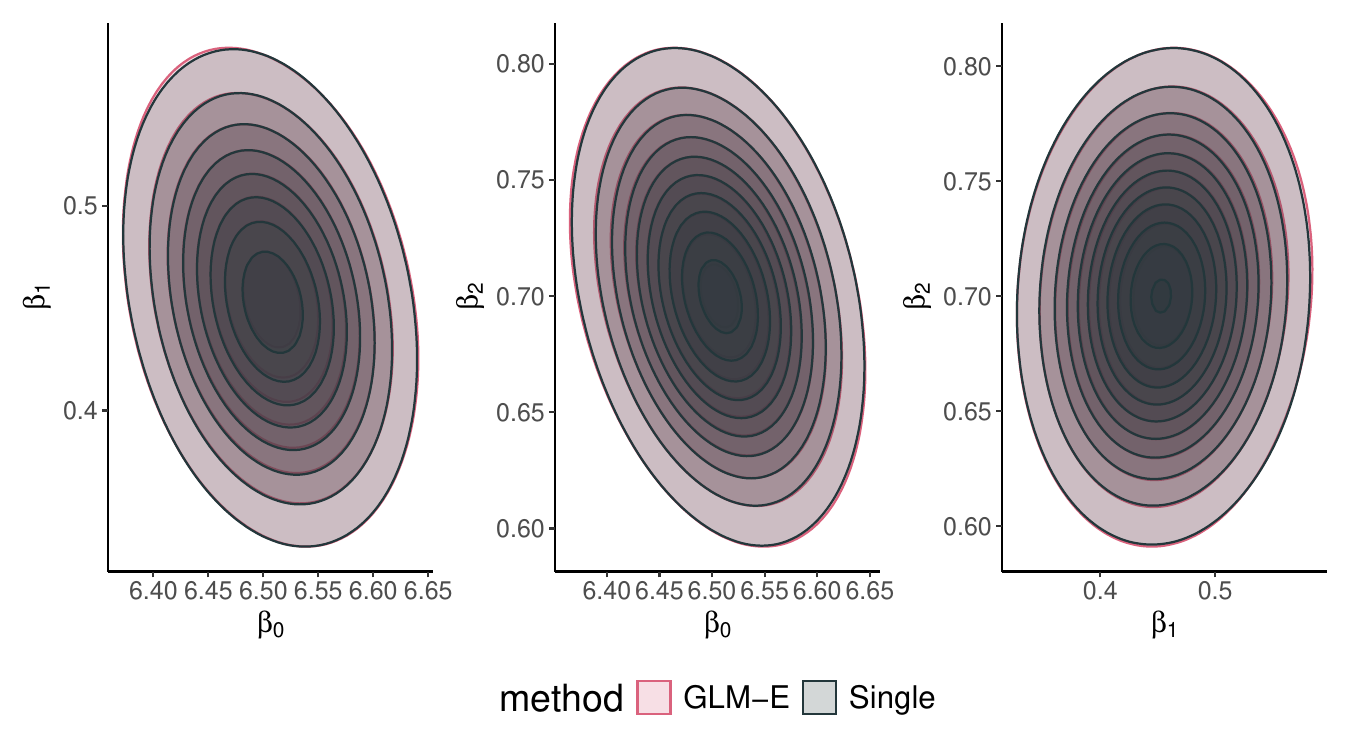}
    \caption{Comparison of bivariate joint posterior distributions for each pair of model parameters. Contours represent kernel density estimates of the joint posterior distributions. The GLM-E contours (pink) are overlaid on top of the single-stage contours (gray).}
\end{figure}

\begin{table}[H]
    \centering
    \def\arraystretch{1.2}
    \begin{tabular}{c c c c}
        \toprule
         & $\beta_0$ & $\beta_1$ & $\beta_2$ \\
         \hline
         Single-stage & 0.170 & 0.167 & 0.146 \\
         PG & 0.038 & 0.032 & 0.045 \\
         HMC & 0.017 & 0.015 & 0.020 \\
         GLM-E & 0.001 & 0.002 & 0.002 \\
         GLM-A & 0.001 & 0.001 & 0.002 \\
         \bottomrule
    \end{tabular}
    \caption{Seconds per effective sample for each coefficient and method. Lower values indicate more efficient sampling.}
\end{table}

\begin{figure}[H]
    \centering
    \includegraphics[width=0.4\linewidth]{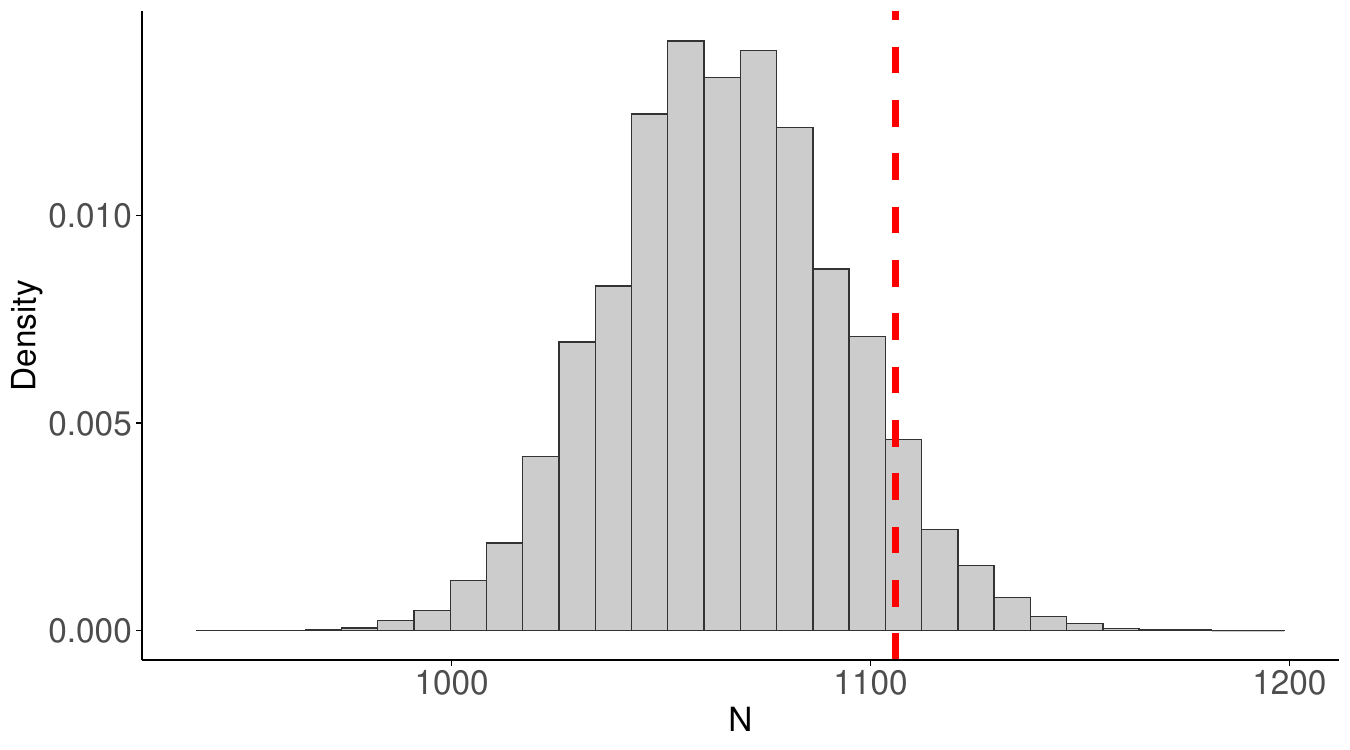}
    \caption{Posterior predictive distribution for $N$ using posterior realizations obtained using the GLM-E method. The true value of $N$ is marked in dashed red.}
\end{figure}

\subsection*{Large windows, point pattern 1}

\begin{figure}[H]
    \centering
    \subfloat[\centering]{{\includegraphics[width=0.45\linewidth]{figs/sim3_intensity.png} }}%
    \qquad
    \subfloat[\centering]{{\includegraphics[width=0.4\linewidth]{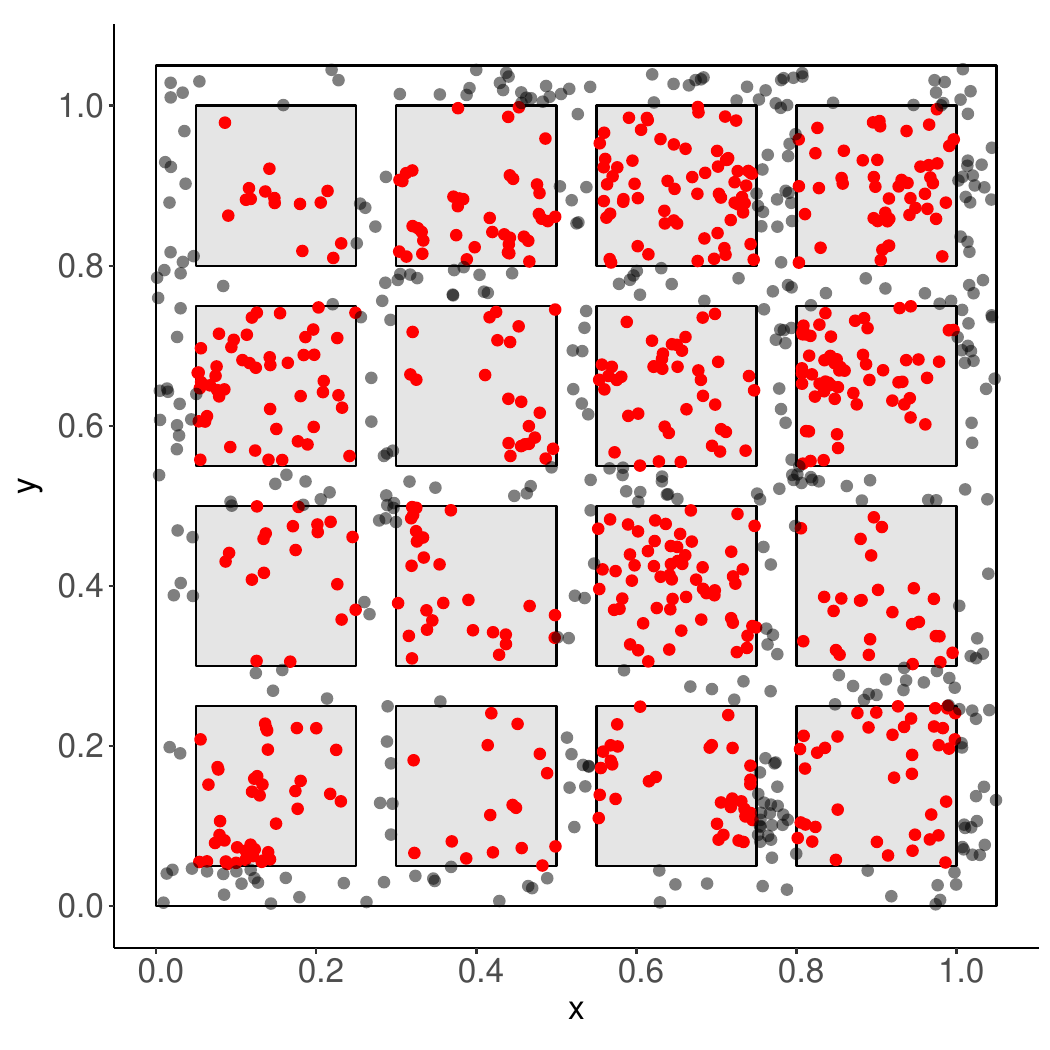} }}%
    \caption{(a) Intensity heat map for simulated data. (b) Realization of simulated points. Points located within the compact windows (red; $n$ = 528) were used to fit the model.}%
\end{figure}

\begin{figure}[H]
    \centering
    \includegraphics[width=0.8\linewidth]{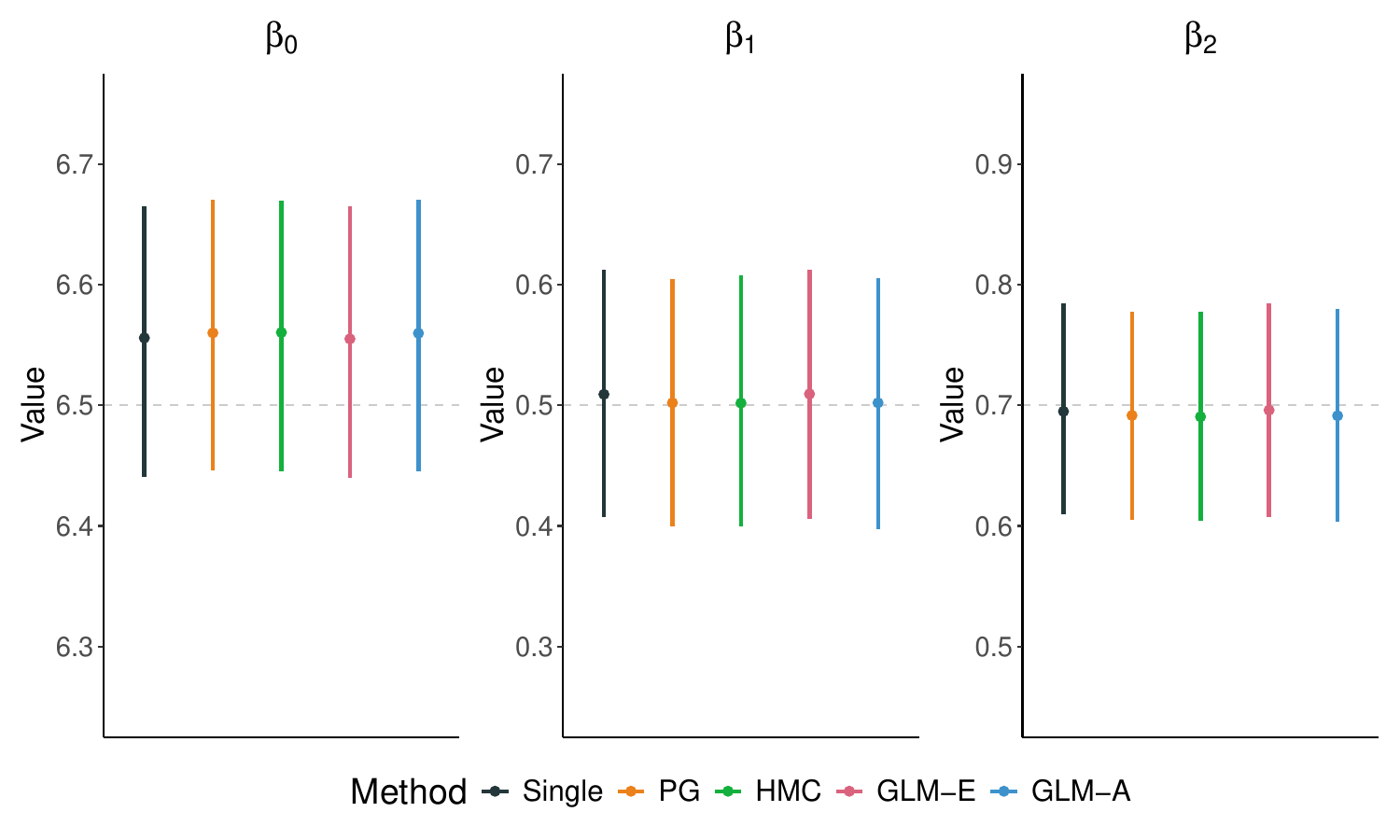}
    \caption{Comparison of marginal posterior 95\% credible intervals for the single-stage method and various multi-stage methods. The true values of the parameters are shown in dashed gray.}
\end{figure}

\begin{figure}[H]
    \centering
    \includegraphics[width=0.8\linewidth]{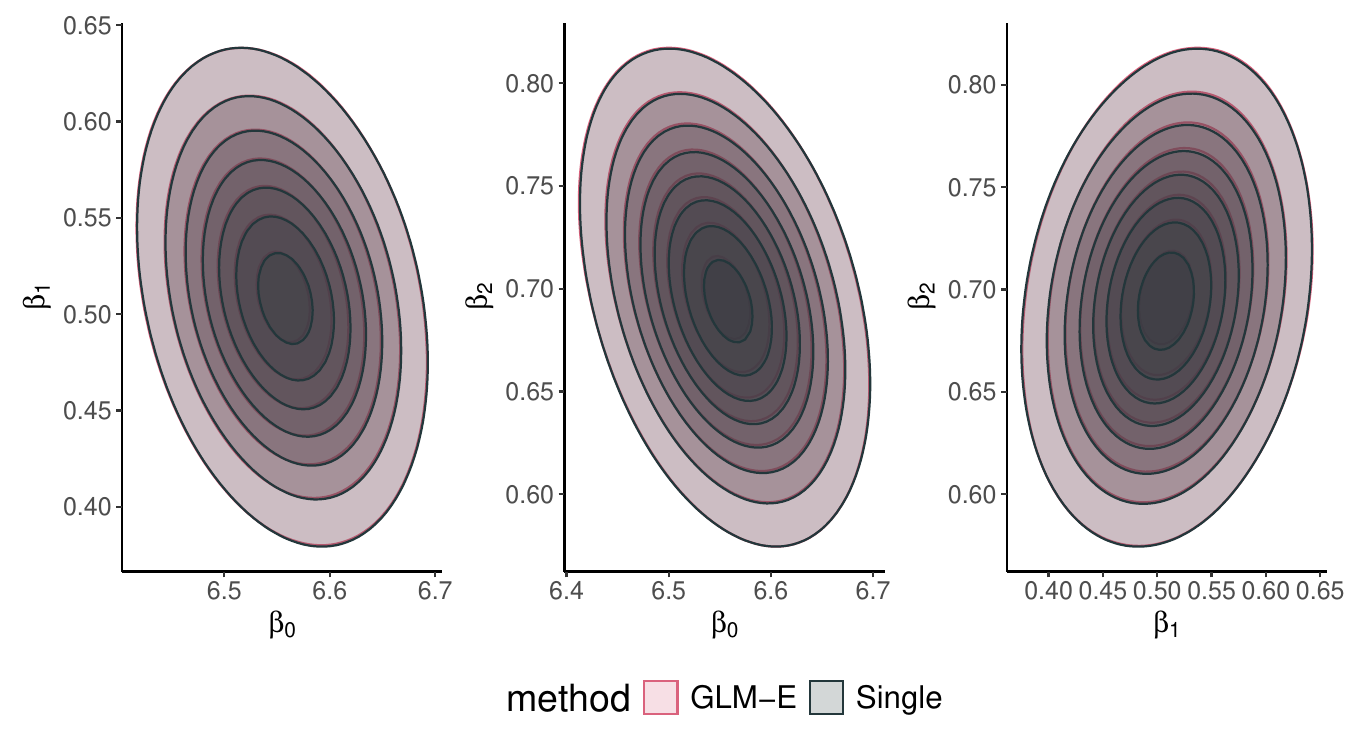}
    \caption{Comparison of bivariate joint posterior distributions for each pair of model parameters. Contours represent kernel density estimates of the joint posterior distributions. The GLM-E contours (pink) are overlaid on top of the single-stage contours (gray).}
\end{figure}

\begin{table}[H]
    \centering
    \def\arraystretch{1.2}
    \begin{tabular}{c c c c}
        \toprule
         & $\beta_0$ & $\beta_1$ & $\beta_2$ \\
         \hline
         Single-stage & 0.128 & 0.139 & 0.137 \\
         PG & 0.028 & 0.029 & 0.035 \\
         HMC & 0.015 & 0.015 & 0.018 \\
         GLM-E & 0.001 & 0.001 & 0.001 \\
         GLM-A & 0.001 & 0.001 & 0.001 \\
         \bottomrule
    \end{tabular}
    \caption{Seconds per effective sample for each coefficient and method. Lower values indicate more efficient sampling.}
\end{table}

\begin{figure}[H]
    \centering
    \includegraphics[width=0.4\linewidth]{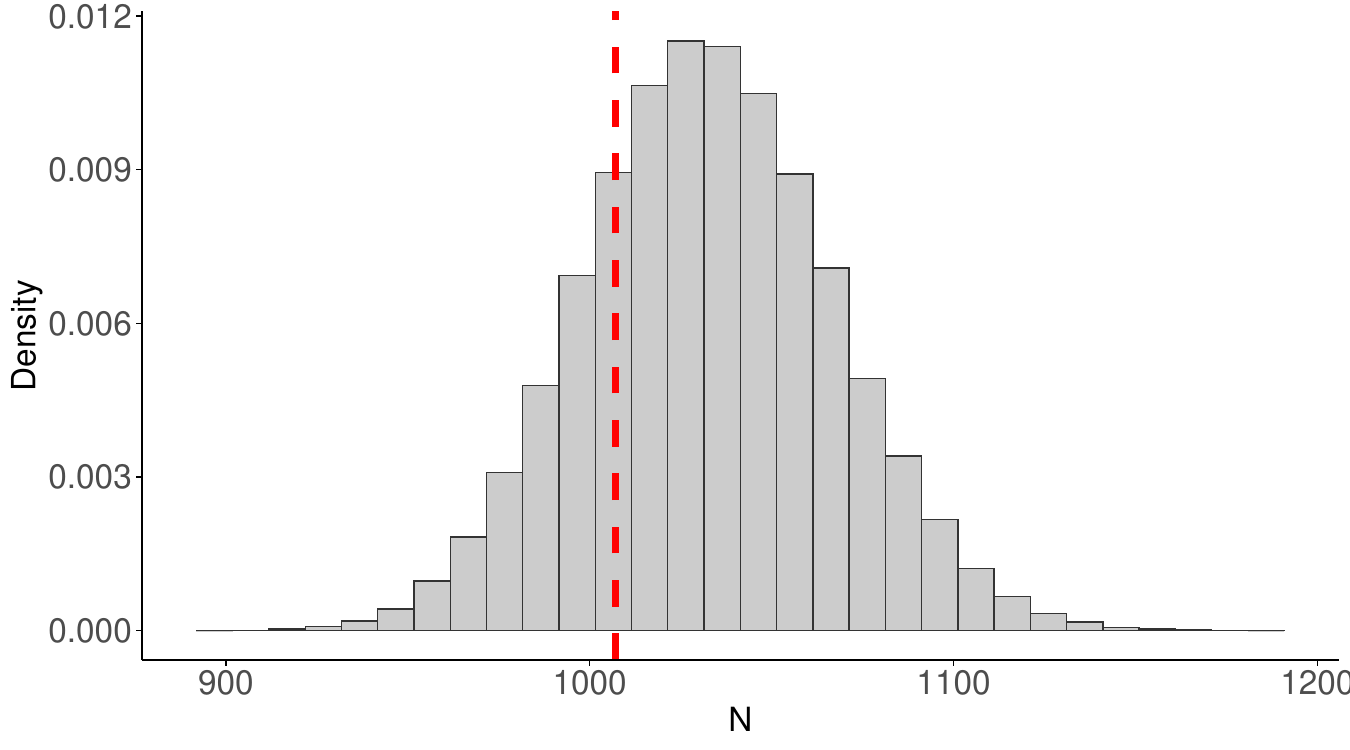}
    \caption{Posterior predictive distribution for $N$ using posterior realizations obtained using the GLM-E method. The true value of $N$ is marked in dashed red.}
\end{figure}

\subsection*{Large windows, point pattern 2}

\begin{figure}[H]
    \centering
    \subfloat[\centering]{{\includegraphics[width=0.45\linewidth]{figs/sim2_intensity.png} }}%
    \qquad
    \subfloat[\centering]{{\includegraphics[width=0.4\linewidth]{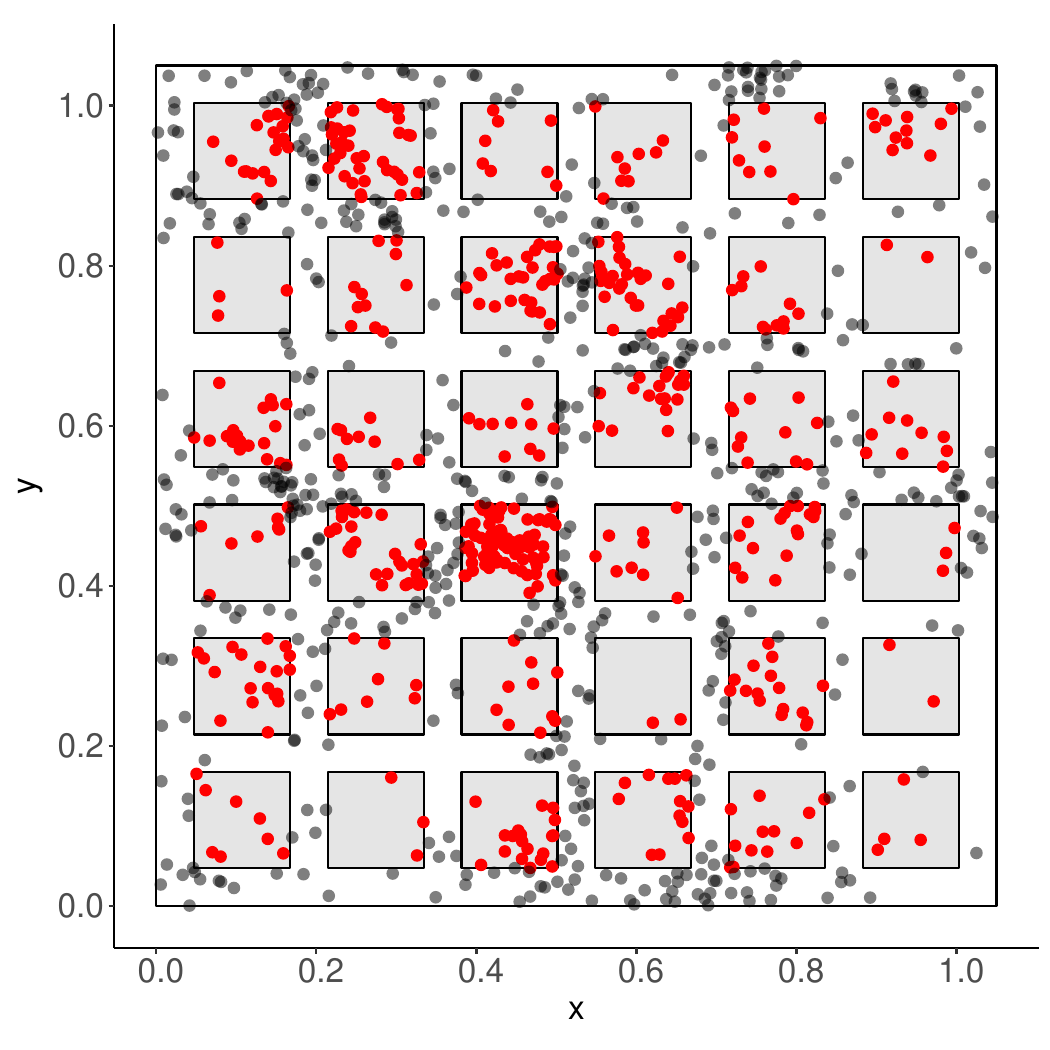} }}%
    \caption{(a) Intensity heat map for simulated data. (b) Realization of simulated points. Points located within the compact windows (red; $n$ = 483) were used to fit the model.}%
\end{figure}

\begin{figure}[H]
    \centering
    \includegraphics[width=0.8\linewidth]{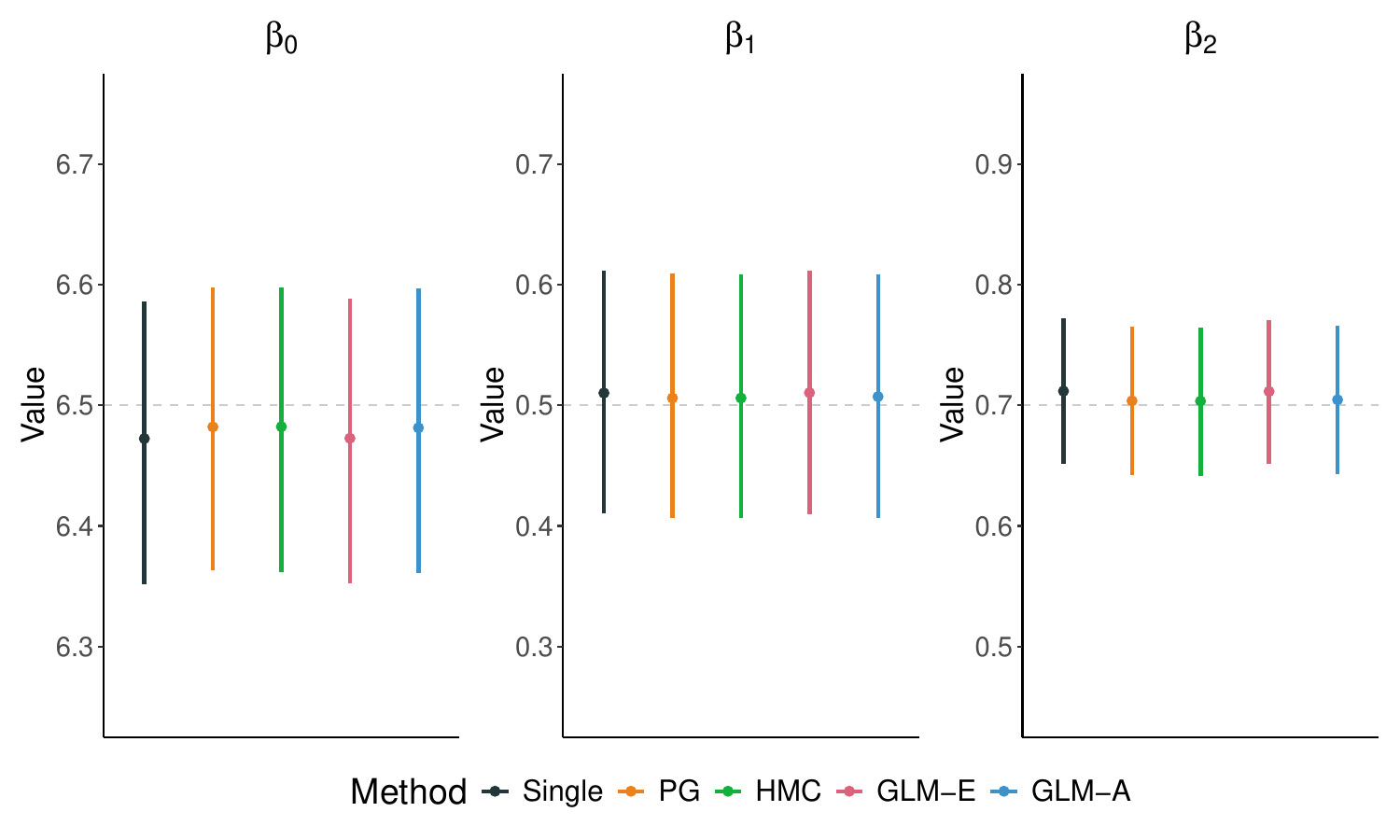}
    \caption{Comparison of marginal posterior 95\% credible intervals for the single-stage method and various multi-stage methods. The true values of the parameters are shown in dashed gray.}
\end{figure}

\begin{figure}[H]
    \centering
    \includegraphics[width=0.8\linewidth]{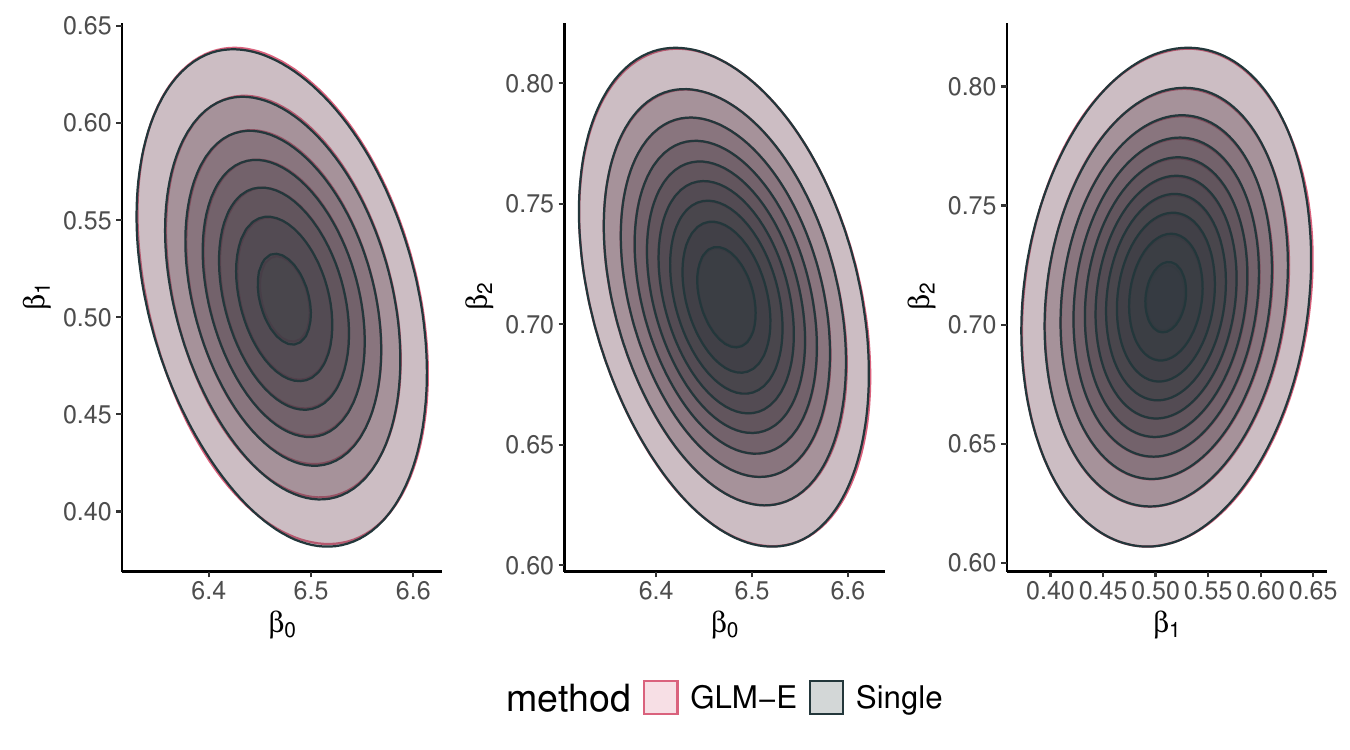}
    \caption{Comparison of bivariate joint posterior distributions for each pair of model parameters. Contours represent kernel density estimates of the joint posterior distributions. The GLM-E contours (pink) are overlaid on top of the single-stage contours (gray).}
\end{figure}

\begin{table}[H]
    \centering
    \def\arraystretch{1.2}
    \begin{tabular}{c c c c}
        \toprule
         & $\beta_0$ & $\beta_1$ & $\beta_2$ \\
         \hline
         Single-stage & 0.162 & 0.174 & 0.157 \\
         PG & 0.040 & 0.037 & 0.048 \\
         HMC & 0.023 & 0.021 & 0.027 \\
         GLM-E & 0.002 & 0.002 & 0.002 \\
         GLM-A & 0.001 & 0.001 & 0.001 \\
         \bottomrule
    \end{tabular}
    \caption{Seconds per effective sample for each coefficient and method. Lower values indicate more efficient sampling.}
\end{table}

\begin{figure}[H]
    \centering
    \includegraphics[width=0.4\linewidth]{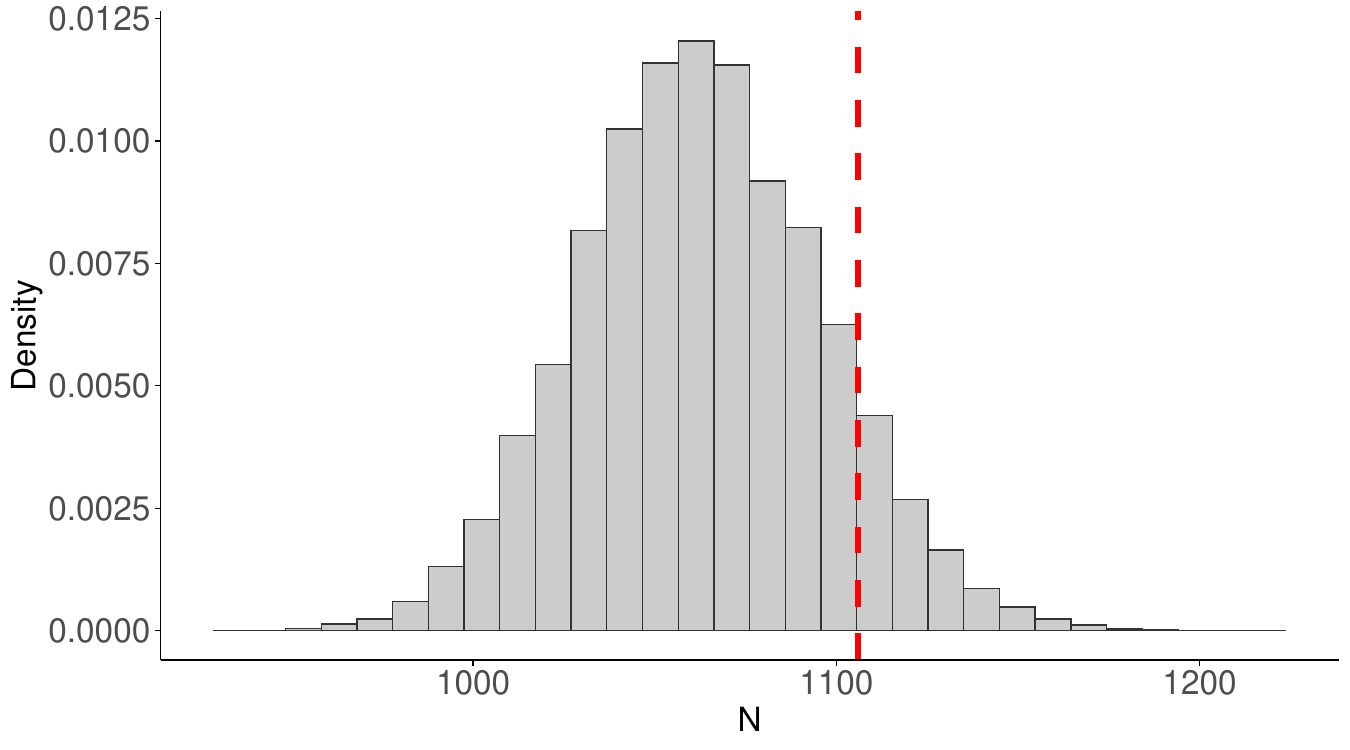}
    \caption{Posterior predictive distribution for $N$ using posterior realizations obtained using the GLM-E method. The true value of $N$ is marked in dashed red.}
\end{figure}

\section*{Appendix C}

We used three covariates (bathymetry, ice proportion, and distance to glacier) to fit the IPP model to harbor seal pup data on 21 June 2007. For the bathymetry covariate, we downloaded a raster with cell size 1/3 arc-second ($\approx$ 10 m) from the National Oceanic and Atmospheric Administration (NOAA) \href{https://www.ncei.noaa.gov/products/nos-hydrographic-survey}{Hydrographic Survey Metadata Database (HSMD)}. The raster values are relative to Mean Lower Low Water, with more negative values corresponding to deeper waters. We used the bathymetry raster grid to construct the remaining covariate rasters. Ice data were collected using the same aerial imagery survey described in section 5. \citet{Kaluzienski2023} conducted an ice segmentation approach to obtain iceberg outlines within each image footprint. We used the iceberg outlines to compute iceberg proportions and performed kriging to complete the ice proportion raster (Appendix D). To construct the the distance to glacier terminus raster, we computed the Euclidean distance from each grid center to the closest point on the glacier terminus.

\section*{Appendix D}

The goals for utilizing the aerial ice imagery were to (1) quantify ice availability from the raw images by creating a rasterized map of ice proportions and (2) produce a map that has the same resolution and spatial extent as the other covariates used in the model (i.e., bathymetry and distance from glacier terminus) across the entire study domain. To ensure an appropriate grid to account for the varying locations of observed ice proportions, the geometries of each image footprint in the survey date were first intersected with a pre-specified raster (Figure 9A). The resulting output was a series of grid cell geometries and their corresponding centroid coordinates. 

Using these grid cell geometries, as well as a binarized version of the observed icebergs, an iterative masking method was used to find the proportion of ice within each grid cell. For each image in the survey date, the raster grid was superimposed onto the image and a logical mask was used to iterate through the grid cell and image intersections. Within a grid cell, the total number of iceberg pixels was divided by the total number of pixels in the grid cell, resulting in an ice proportion value. This process was repeated for each grid cell in an image, and for each image in the survey date (Figure 22B). Because ice was only observed within the areas of the image footprints, the calculated ice proportions and the \texttt{LatticeKrig R} package \citep{Nychka2023} were used to produce an interpolated surface of ice proportions across the entire survey domain. After performing kriging, a raster was produced, where ice proportion estimates were available at the same resolution as the other covariates used, and across the entire survey domain. 

\begin{figure}
    \centering
    \includegraphics[width=\linewidth]{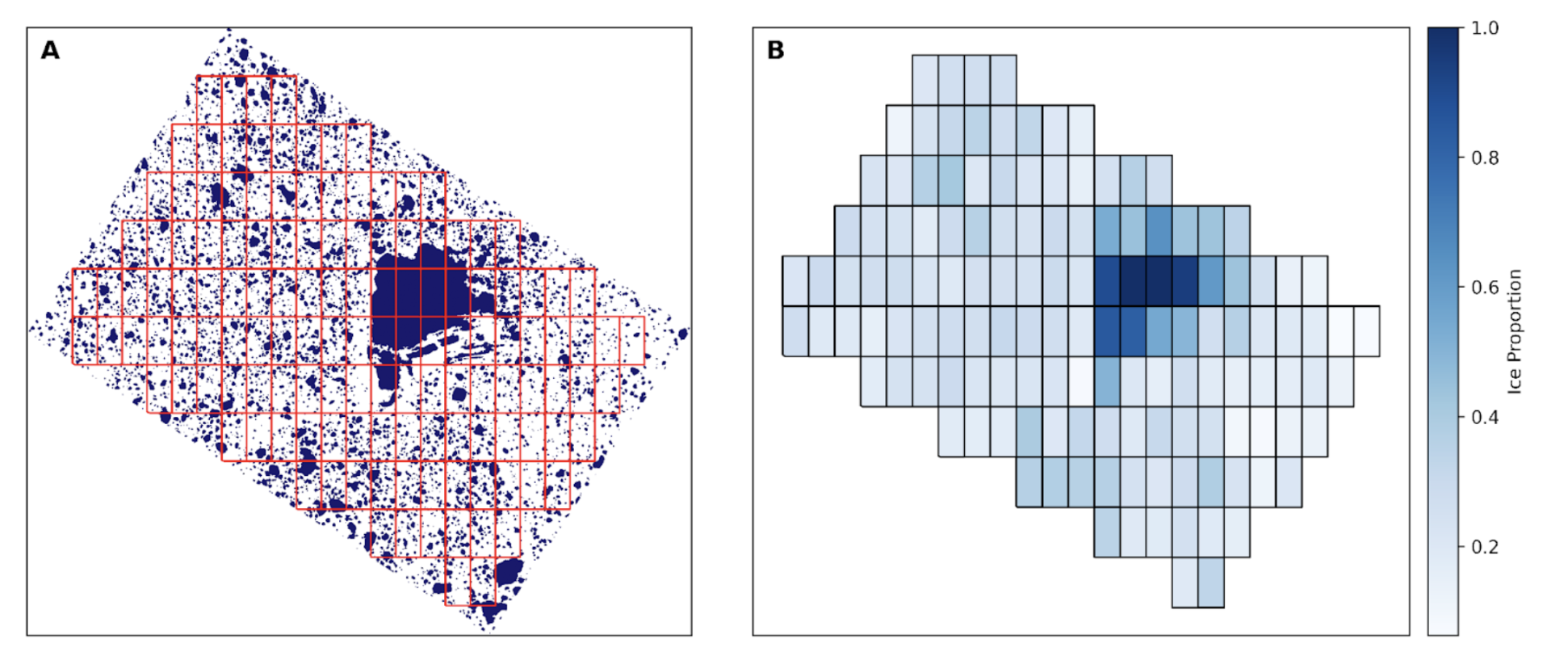}
    \caption{A) Pre-processed iceberg outlines for one image footprint in the 21 June 2007 survey. B) Computed ice proportions on the pre-specified covariate grid.}
\end{figure} 

\section*{Appendix E}

To check whether the observed data reasonably arises from the fitted IPP model, we performed a posterior predictive check based on the L-function, a function used to summarize the level of clustering in a spatial point pattern at varying distances. For each $\beta_0^{(k)}$ and $\boldsymbol{\beta}^{(k)}$ posterior realization obtained from the multi-stage GLM-E method, we simulated points within the image footprints and computed the corresponding L-function using an isotropic edge correction. These were then compared to the L-function computed for the observed harbor seal pup locations. 

We performed this model checking technique for the model with and without the NN basis expansion. For the model without the NN basis expansion, we found that the L-function computed from the observed pup data did not lie within the realm of posterior predictive L-functions (Figure 23), indicating that the model did not sufficiently account for clustering in the observed point pattern. However, after refitting the model with the inclusion of the NN basis expansion, we found that the L-function computed from the observed pup data lie within the realm of posterior predictive L-functions (Figure 24), indicating a reasonable model fit. 

\begin{figure}[h!]
    \centering
    \includegraphics[width=0.5\linewidth]{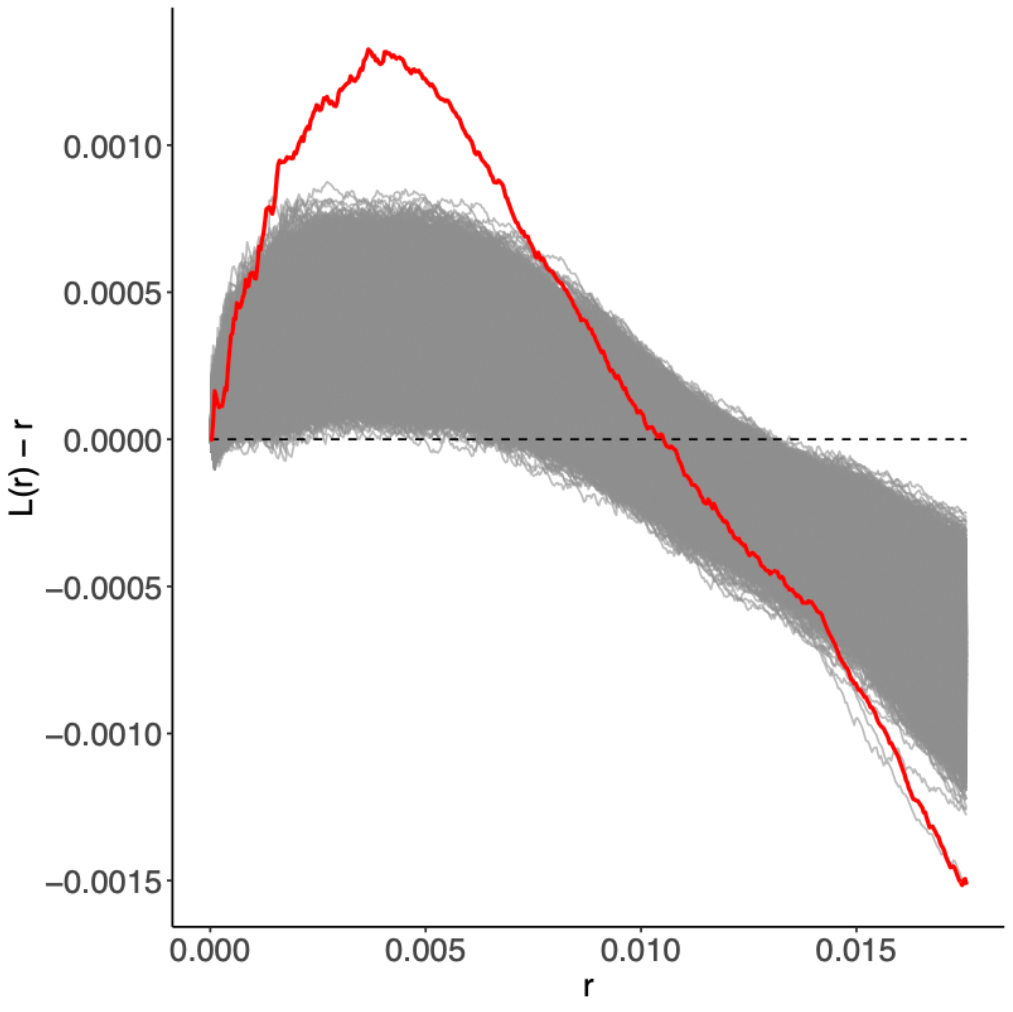}
    \caption{Posterior predictive L-functions from the model without NN basis expansion (gray) compared to the observed L-function (red).}
\end{figure}

\begin{figure}[h!]
    \centering
    \includegraphics[width=0.5\linewidth]{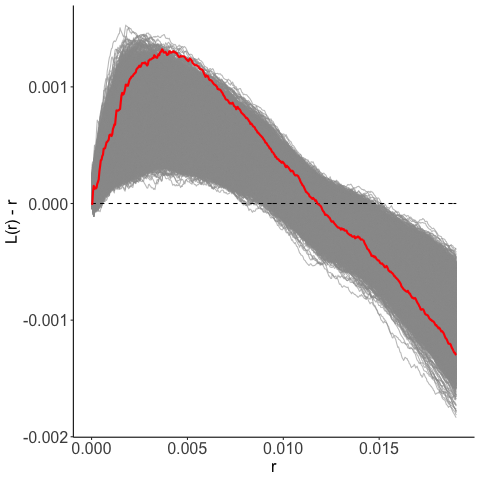}
    \caption{Posterior predictive L-functions from the model with NN basis expansion (gray) compared to the observed L-function (red).}
\end{figure}

\end{document}